%% file: icgt2023.tex
\documentclass[runningheads,envcountsect]{llncs}

\usepackage{pdfpages}

\usepackage[T1]{fontenc}

\usepackage[colorlinks]{hyperref}
\usepackage[links,references]{agda}
\usepackage{amsmath}  
\usepackage{amssymb}  
\usepackage{andreamacro}
\usepackage{array}
\usepackage{booktabs}
\usepackage{capt-of}
\usepackage{comment}
\usepackage{realboxes}
\usepackage{enumitem} 
\usepackage{float} 
\usepackage{graphicx} 
\usepackage{imakeidx} 
\usepackage{mathrsfs}
\usepackage{mathtools} 
\usepackage{setspace} 
\usepackage{stmaryrd} 
\usepackage{tabularx}
\usepackage{tikz-cd}
\usepackage{tikz}
\usepackage{tikzit}
\usepackage{xcolor}
\usepackage{xspace}
\usetikzlibrary{shapes.geometric}
\usetikzlibrary{positioning}

\makeatletter
\def\defthm#1#2{%
  \newtheorem{#1}{#2}[section]%
  \expandafter\def\csname #1autorefname\endcsname{#2}%
  \expandafter\let\csname c@#1\endcsname\c@theorem}
\makeatother
\defthm{proposizia}{Proposition}

\usepackage{etoolbox}
\makeatletter
\patchcmd{\hyper@makecurrent}{%
    \ifx\Hy@param\Hy@chapterstring
        \let\Hy@param\Hy@chapapp
    \fi
}{%
    \iftoggle{inappendix}{
        \@checkappendixparam{chapter}%
        \@checkappendixparam{section}%
        \@checkappendixparam{subsection}%
        \@checkappendixparam{subsubsection}%
        \@checkappendixparam{paragraph}%
        \@checkappendixparam{subparagraph}%
    }{}%
}{}{\errmessage{failed to patch}}

\newcommand*{\@checkappendixparam}[1]{%
    \def\@checkappendixparamtmp{#1}%
    \ifx\Hy@param\@checkappendixparamtmp
        \let\Hy@param\Hy@appendixstring
    \fi
}
\makeatletter

\newtoggle{inappendix}
\togglefalse{inappendix}

\apptocmd{\appendix}{\toggletrue{inappendix}}{}{\errmessage{failed to patch}}

\definecolor{pisa}{RGB}{0, 100, 150}
\definecolor{verdino}{RGB}{0, 140, 140}
\definecolor{colore_review}{RGB}{242, 96, 53}
\definecolor{bluino}{RGB}{0, 0, 255}
\definecolor{colore_red}{RGB}{230, 31, 31}
\definecolor{colore_blue}{RGB}{54, 133, 236}
\definecolor{colore_purple}{RGB}{181, 71, 141}
\usepackage{todonotes}

\input{work.tikzstyles}
\input{work-definitions.tex}

\usepackage{newunicodechar}
\usepackage{relsize}
\input{symbols}

\newcommand{\agda}[2]{(\href{#1}{\texttt{\hspace{0.07em}\raisebox{-0.2em}{\includegraphics[height=1em]{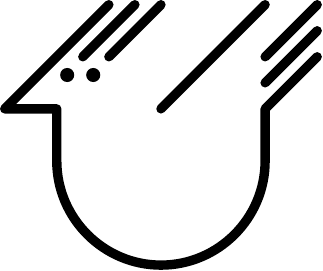}}\hspace{0.20em}\AgdaFontStyle{#2}}})}

\def\reviewcolor{}

\begin{document}

\title{Specification and verification of a linear-time temporal logic for graph transformation\thanks{Research partially supported by the Italian MIUR project
PRIN 2017FTXR7S ``IT-MaTTerS'' and by the University of Pisa project PRA\_2022\_99 ``FM4HD''.}}

\author{Fabio Gadducci \and Andrea Laretto \and Davide Trotta}

\institute{
  Department of Computer Science, University of Pisa,
  Italy\\
  \email{fabio.gadducci@unipi.it},
  \email{trottadavide92@gmail.com}
\\
    Department of Software Science, Tallinn University of Technology, Estonia
    \email{andrea.laretto@taltech.ee}
 }

\authorrunning{Gadducci, Laretto and Trotta}

\titlerunning{Specification and verification of a linear-time logic for graph transformation}
\maketitle%
\begin{abstract}%
We present a first-order linear-time temporal logic for reasoning about the evolution of directed graphs.
Its semantics is based on the counterpart paradigm, thus allowing our logic to represent the creation, duplication, merging, and deletion of elements of a graph
as well as how its topology changes over time. We then introduce a positive normal forms presentation, thus simplifying the actual process
of verification. We provide the syntax and  semantics of our logics with a computer-assisted
formalisation using the proof assistant Agda, and we round up the paper by highlighting the crucial aspects of our formalisation and the practical use of quantified
temporal logics in a constructive proof assistant.
\end{abstract}



\section{Introduction}
\label{ch:introduction}

Among the many tools provided by formal methods, temporal logics have proven to be one of the most effective techniques for the verification of both large-scale and stand-alone programs.
Along the years, the research on these logics focused on improving the algorithmic procedures for the verification process as well as on finding sufficiently expressive fragments of these logics for the specification of complex multi-component systems.
Several models for temporal logics have been developed, with the leading example being transition systems, also known as Kripke structures. In a transition system, each state represents a configuration of the system and each transition identifies a possible state evolution. Often one is interested in enriching the states and transitions given by the model with more structure, for example by taking states as algebras and transitions as algebra homomorphisms. A prominent use case is that of graph logics~\cite{the-monadic-second-order-logic-of-graphs-I:1990,the-monadic-second-order-logic-of-graphs-XII:2000,expressiveness-complexity-graph-logic:2007}, where states are specialised as graphs and transitions are families of (partial) graph morphisms. These logics combine temporal and spatial reasoning and allow to express the possible transformation of a graph topology over time.

\noindent\textbf{Quantified temporal logics.} Under usual temporal logics, such as \LTL and \CTL~\cite{temporal-and-modal-logic:1990}, the states of the model are taken as atomic, with propositions holding for entire states: on the other hand, one of the defining characteristics of  graph logics is that they permit to reason and to express properties on the individual elements of the graph. Despite their undecidability~\cite{fixpoint-extensions-temporal-description-logics:2003,monodic-fragments-first-order-temporal-logics:2001}, quantified temporal logics have been advocated in this setting due to their expressiveness and the possibility for quantification to range over the elements in the states of the model.

The semantical models of these logics require some ingenuity, though. Consider a simple model with two states $s_0,s_1$, two transitions $s_0 \to s_1$ and $s_1 \to s_0$, and an item $i$ that appears only in $s_0$. Is the item $i$ being destroyed and recreated, or is it just an identifier being reused multiple times? This issue is denoted in the literature as the \emph{trans-world identity problem}~\cite{counterpart-theoretic-semantics-for-modal-logic:1979,quantified-modal-logic-ontology-physical-objects:2006}. {\reviewcolor A solution consists in fixing a single set of universal items, which gives identity to each individual appearing in the states of the model. Since each item $i$ belongs to this universal domain, it is exactly the same individual after every temporal evolution in $s_1$.} However, this means that transitions basically behave as injections among the items of the states, and this view is conceptually difficult to reconcile with the simple model sketched above where we describe the destruction and recreation of a given item.
{\reviewcolor Similarly, the possibility of cloning items is then ruled out, since it is impossible to accomodate it with the idea of evolution steps as injections.}

\noindent\textbf{Counterpart semantics.} A solution to this problem was proposed by Lewis~\cite{counterpart-theory-quantified-modal-logic:1968} with the \emph{counterpart paradigm}: instead of a universal set of items, each state identifies a local set of elements, and (possibly partial) morphisms connect them by carrying elements from one state to the other. This allows us to speak formally about entities that are destroyed, duplicated, (re)created, or merged, and to adequately deal with the identity problem of individuals between graphs.

In \cite{counterpart-semantics-second-order-mu-calculus:2012}, a counterpart-based semantics is used to introduce a set-theoretical semantics of a $\mu$-calculus with second-order quantifiers. This modal logic provides a formalism that enriches states with algebras and transitions with partial homomorphisms, subsuming the case of graph logics.
These models are generalised to a categorical setting in \cite{presheaf-semantics-qtl:2021} by means of relational presheaves, building on the ideas presented in \cite{modal-tense-predicate-logic-models-in-presheaves-categorical-conceptualization:1988,relational-partial-variable-sets-basic-predicate-logic:1996}. The models are represented with categories and (families of) relational presheaves, which give a categorical representation for the states-as-algebras approach with partial homomorphisms. The  temporal advancement of a system is captured by equipping categories with the notion of  \emph{one-step} arrows for a model, and the categorical framework is then used to introduce a second-order linear temporal logic \QLTL.

\noindent\textbf{Classical semantics and positive normal form.} We start by introducing the notion of counterpart models in \autoref{sec:counterpart_semantics}, and present the (admittedly rather straightforward) syntax of our temporal logic \QLTL as well as its counterpart-based semantics in \autoref{ch:qltl_classical}, using a standard set-theoretic perspective, with satisfiability given inductively as a logical predicate. Unlike \cite{counterpart-semantics-second-order-mu-calculus:2012,presheaf-semantics-qtl:2021} where the models use \emph{partial functions}, we generalise to \emph{relations}, thus modelling the duplication of elements by allowing it to have multiple counterparts.
In \autoref{sec:pnf} we present some results on the positive normal forms, where the models may use either partial morphisms or relations, and we highlight their differences. Positive normal forms (i.e., where negation is defined only for atomic formulae) are a standard tool of temporal logics, since they simplify its theoretical treatment as well as facilitating model checking algorithms~\cite{tableau-finite-ltl:2022,regular-vacuity:2005}. {\reviewcolor The use of relations instead of (possibly partial) functions weaken the expressiveness
of such normal forms, and require the introduction of additional operators for the logics. However, the duplication of individuals is a central feature of graph transformation formalisms such as Sequi-Pushout~\cite{sesquipushout}, and thus worthy of investigation.}

\noindent\textbf{Temporal logics in Agda.}
An additional contribution of our work is a computer-assisted formalisation using the dependently typed proof assistant Agda \cite{dependently-typed-programming-agda:2009} of the models, semantics, and positive normal forms of the logic presented in this paper.
We introduce the main aspects of the mechanisation in \autoref{ch:agda_formalisation}, which can be adapted for counterpart-based models whose worlds are
algebras on any multi-sorted signature, even if for the sake of presentation in this paper we restrict our attention to graph signatures.
%
A formal presentation of a temporal logic in a proof assistant has several advantages: it solidifies the correctness and coherence of the mathematical ideas presented in the work, as they can be independently inspected and verified concretely by means of a software tool;
moreover, the mechanisation effectively provides a playground in which the mechanisms and validity of these logics can be expressed, tested, and experimented with.

To the best of our knowledge, few formalisations of temporal logics have been provided with a proof assistant, and none of these comes equipped with a counterpart-based
semantics.
This work constitutes a step towards the machine-verified use of temporal logics by embedding in an interactive proof assistant a
 quantified extension of \LTL that can reason on individual elements of states.

\section{Counterpart Models}
\label{ch:qltl_classical}
\label{sec:counterpart_semantics}

This section introduces our models for system evolution.
We consider the instantiation of counterpart models to the case where each world is not associated to a mere (structureless)
set of individuals, but to a directed graph with its evolution in time being represented by suitable relations preserving the graph structure.

\begin{definition}\label{def:graph}
    A \textbf{(directed) graph} is a 4-tuple $G := \abr{N,E,s,t}$ such that
        $N$ is a set of nodes,
        $E$ is a set of edges, and
        $s,t : E \to N$ are two functions assigning a \emph{source} node $s(e) \in N$ and a \emph{target} node $t(e) \in N$ to each edge $e \in E$, respectively. The set of all directed graphs is denoted as $\textsf{Graphs}$.
\end{definition}

\begin{definition}\label{def:graph_relational_homomorphism}
    A \textbf{graph (relational) morphism} between two graphs $G = \abr{N,E,s,t}$ and $G' = \abr{N',E',s',t'}$ is a pair $R := \abr{R_N,R_E}$ such that
        $R_N \subseteq N \times N'$ and $R_E \subseteq E \times E'$ are relations between nodes and edges of the graphs such that $e_1 R_E e_2$ implies $s(e_1) R_N s'(e_2)$ and $t(e_1) R_N t'(e_2)$.
        Given  graphs $G,G'$, the set of graph morphisms is denoted $\textsf{GraphRel}(G,G') \subseteq \mathscr{P} ((N \times N') \times (E \times E'))$.
\end{definition}

\begin{definition}\label{def:counterpart_model}
    A \textbf{counterpart model} is a triple $\mathfrak{M} := \abr{W,D,\C}$ such that
    \begin{itemize}
        \item $W$ is a non-empty set of elements, called \emph{worlds},
        \item $D : W \to \textsf{Graphs}$ is a function assigning a directed graph to each world,
        \item $\C : W \times W \to \mathscr{P}(\textsf{GraphRel}(D(\omega),D(\omega')))$ is a function assigning to every pair $\abr{\w,\w'}$ a set of graph
         morphisms $\C \abr{\w,\w'} \subseteq \textsf{GraphRel}(D(\w),D(\w'))$, where every $C \in \C \abr{\w,\w'}$ is a graph morphism between
         the graphs associated to the two worlds. We refer to these as \textbf{atomic counterpart relations}.
    \end{itemize}
\end{definition}

Given two worlds $\w$ and $\w'$, the set $\C \abr{\w,\w'}$ is the collection of atomic transitions from $\w$ to $\w'$, defining the possible ways we can access worlds with a \emph{one-step transition} in the system. When the set $\C \abr{\w,\w'}$ is empty, there are no atomic transitions from $\w$ to $\w'$.
Each atomic relation $C \in \C \abr{\w,\w'}$ connects the nodes and edges between two worlds $\w$ and $\w'$, intuitively identifying them as the same component
after one time evolution of the model. For example, if we consider two nodes $n\in D(\w)_N$ and $n'\in D(\w')_N$ and a relation $C \in \C \abr{\w,\w'}$, if $\abr{n, n'} \in C_N$ then $n'$ represents a future development of the node $n$ via $C$.

\begin{definition}
    A node $s' \in D(\w')_N$ \textbf{is the counterpart of} $s \in D(\w)_N$ through a counterpart relation $C$ whenever $\abr{s,s'} \in C_N$,
    and similarly for edges.
\end{definition}

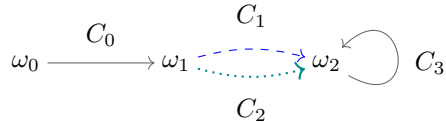
\begin{figure}[H]
    \centering
    \vspace{-1.3em}
    \input{algebraic_counterpart_w.tikz}
    \caption{Graphical representation of the worlds and accessibility relations of a model.}
    \vspace{-1.3em}
    \label{fig:counterpart_w}
\end{figure}

\begin{figure}[H]
    \centering
    \input{running_example_graph.tikz}
    \vspace{-2.0em}
    \caption{Graphical representation of a model.}
    \vspace{-1.3em}
    \label{fig:counterpart_w_model_example}
\end{figure}
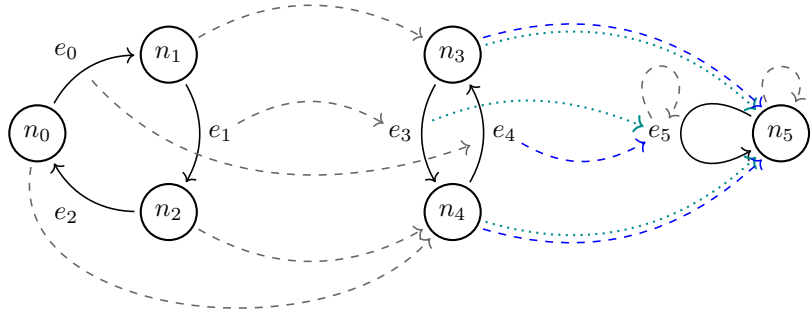

\begin{example}[Counterpart model]
    \label{ex:counterpart_model}
We give an example of a counterpart model
by indicating the set of worlds $\set{\omega_0,\omega_1,\omega_2}$ and the cardinality of the sets of relations $\C\abr{\omega,\omega'}$ in \autoref{fig:counterpart_w}; the graph structures associated to each world and the graph morphisms connecting them are shown in \autoref{fig:counterpart_w_model_example}. There are two counterpart relations $C_1,C_2 \in \C\abr{\omega_1,\omega_2}$ between $\omega_1$ and $\omega_2$, and we use \textcolor{bluino}{blue dashed} and \textcolor{verdino}{green dotted} lines to distinguish $C_1$ and $C_2$, respectively. \autoref{graphexample} includes an explicit description of the model and its counterpart relations.
\end{example}

The use of \emph{relations} as transitions allows us to model the removal of edges and nodes of a graph, by having no counterpart for them in the next state.
For example, if there is no edge $e'\in D(\w')_E$ such that $\abr{e,e'} \in C_E$, we conclude that the edge $e$ has been removed by $C$.
Similarly, the duplication of a node is represented by connecting it with two instances of the counterpart relation, e.g. by having elements $n'_1,n'_2 \in D(\w')_N$
such that $\abr{n,n'_1} \in C_N$ and $\abr{n,n'_2} \in C_N$.

{\reviewcolor
The simple counterpart model in \autoref{ex:counterpart_model} displays the effects of merging and deletion.
The first counterpart relation $C_0$ merges the nodes $n_0$ and $n_2$ of $\omega_0$, yet this act does not generate a cycle: in fact,
$e_2$  is deleted, since it is not connected to a counterpart in $\omega_1$.
Similarly, both counterpart relations $C_1$ and $C_2$ merge the nodes $n_3$ and $n_4$ of $\omega_1$, while they differ in which edge they remove when transitioning from $\omega_1$ to $\omega_2$, by deleting either the edge $e_3$ or $e_4$, respectively. Note that in both cases the nodes $n_3$ and $n_4$ need to be preserved, albeit possibly merged, in order to ensure that the relations considered are actually (relational) morphisms of graphs.}

\subsection{Counterpart relations and traces}

We assume hereafter a fixed counterpart model $\mathfrak{M} := \abr{W,D,\C}$
%
and we formally introduce the idea behind counterpart relations.
We indicate composition of graph morphisms in diagrammatic order: as an example, given $C \in \textsf{GraphRel}(G_1, G_2)$ and $C' \in \textsf{GraphRel}(G_2, G_3)$, the composite graph morphism is denoted with $C ; C'  \in \textsf{GraphRel}(G_1, G_3)$ and is such that $(C ; C')_N = \set{(a,c) \mid \exists b \in {G_2}_N.\ \abr{a,b} \in C_N \land \abr{b,c} \in C'_N}$, and similarly for edges.


\begin{definition}\label{def:counterpart_relation}
    A graph morphism $C \in \textsf{GraphRel}(D(\w), D(\w'))$
    is a \textbf{counterpart relation} of the model if one of the following cases holds
    \begin{itemize}
        \item $C$ is the identity graph morphism;
        \item $C \in \C \abr{\w,\w'}$ is an atomic graph morphism given by the model $\mathfrak{M}$;
        \item $C$ is the composite of a sequence of counterpart relations $C_0 ; \cdots ; C_n$ with $C_i \in \C \abr{\w_i,\w_{i+1}}$.
    \end{itemize}
\end{definition}

Note that the composition $C ; C' \in \textsf{GraphRel}(D(\w_1), D(\w_3))$ of two atomic counterpart relations $C\in \C \abr{\w_1,\w_2}$ and $C' \in \C \abr{\w_2,\w_3}$ might not be atomic, {\reviewcolor  and the models define only atomic transitions. Transitioning directly between two graphs might differ from transitioning through an intermediate one, since the direct transition is not necessarily the composition of the two counterpart relations. Moreover, the former requires one evolution step, the latter two.}


As is the case of \LTL where we can identify traces connecting linearly evolving states, see for example~\cite{handbook-modal-logic:2007}, we can consider linear sequences of counterpart relations {\reviewcolor  providing a list of sequentially accessible worlds with associated graphs.}

\begin{definition}\label{def:trace}
    A \textbf{trace} $\sigma$ 
    is an infinite sequence of atomic counterpart relations $(C_0,C_1,\dots)$ such that $C_i \in \C \abr{\w_i,\w_{i+1}}$ for any $i \geq 0$.
\end{definition}

{\reviewcolor In other words, a trace identifies a path in the graph induced by the counterpart model, with words as nodes and atomic counterpart relations as edges.}
Given a trace $\sigma = (C_0,C_1,\dots)$, we use $\sigma_i := (C_{i},C_{i+1},\dots)$ to denote the trace obtained by excluding the first $i$ counterpart relations. We use $\omega_0, \omega_1, \dots$ and $\omega_i$ to indicate the {\reviewcolor worlds} of the trace $\sigma$ whenever it is clear from the context. 
%
Similarly, we denote with $C_{\leq i}$ the composite relation $C_0 ; \cdots ; C_{i-1}$ from the first world $\w_0$ up to the $i$-th world $\w_i$ through the relations given by the trace $\sigma$. Should $i=0$, the relation $C_{\leq 0}$ is the identity graph morphism on $\w_0$.

\section{Quantified linear temporal logic}
\label{ch:qltl_classical}

We present the syntax and semantics of our (first-order) quantified linear temporal logic \QLTL by adopting a standard set-theoretic presentation.

\subsection{Syntax and semantics of \QLTL}

{\reviewcolor
Since free variables may appear inside formulae, we recall the usual presentation of context and terms-in-context.
For the sake of simplicity we consider algebras whose terms represent directed graphs,
as per \autoref{def:graph}.
More precisely, the signature has two sorts \textsf{E} and \textsf{N} and two functions symbols $s$ and $t$, obviously representing the source and target functions
on edges, respectively. We assume a set of sorted variables $\X = \X_N \uplus \X_E$ and
define a typed context $\G$ as a finite subset of such variables,} using the notation $[\G]\ n : \textsf{N}$ to indicate that the term
$n$ has type \textsf{N} and it is constructed in a typed context $\G$, that is, its free variables are contained among those occurring in $\G$
(and similarly for $[\G]\ e : \textsf{E}$).

In order to give a simpler presentation for the semantics of temporal logics, it is customary to exclude the elementary constructs that can be expressed in terms of other operators. {\reviewcolor We thus present \QLTL with a minimal set of standard operators and derive conjunction and universal quantification by using negation.}

\begin{definition}[\QLTL]\label{def:qltl_syntax}
    Let $\Gamma$ be a typed context on the set of variables $\X$. The set $\F^\QLTL_\Gamma$ of \QLTL formulae is generated by the following rules
    \[ \psi := \true
    \mid e_1 =_E e_2
    \mid n_1 =_N n_2 \]
   \[   \phi := \psi
        \mid \neg \phi
        \mid \phi \lor \phi
        \mid \exists_N x.\phi
        \mid \exists_E x.\phi
        \mid \next \phi
        \mid \phi \until \phi
        \mid \phi \wuntil \phi.
    \]
    where $[\G]\ e_i : \textsf{E}$ and $[\G]\ n_i : \textsf{N}$ for $i = 1, 2$.
\end{definition}

The above definition actually provides formulae-in-context: we use the notation $[\G]\phi$ to indicate that a formula $\phi$ belongs to
$\F^\QLTL_\Gamma$ for the typed context $\G$. Clearly, saying that $[\G]\phi$ is the same as stating that
$\fv(\phi) \subseteq \G$, i.e. the free variables of $\phi$ are contained among those occurring in $\G$.
For both terms and formulae we omit the bracketed context whenever it is unnecessary to specify it.

{\reviewcolor The letter $\psi$ denotes the set of
\emph{atomic formulae}, built out of two sorted equivalence predicates.
Given two edge terms $e_1,e_2 : \textsf{E}$, the formula $e_1 =_E e_2$ indicates that the two edges coincide in the graph associated to the current world,
and similarly for two node terms $n_1,n_2 : \textsf{N}$ and the formula $n_1 =_N n_2$.}

\newpage
The existential operators $\exists_N x.\phi$ and $\exists_E x.\phi$ can be used to express the existence of a node (edge, respectively) in the current graph satisfying a certain property $\phi$, where the variable $x$ is allowed to appear as a free variable of $\phi$.

The \emph{next} operator \nextP expresses the fact that a certain property $\phi$ has to be true at the next state.
The \emph{until} operator \untilP indicates that the property $\phi_1$ has to hold at least until the property $\phi_2$ becomes true, which must hold at the present or future time.
Finally, the \emph{weak until} operator \wuntilP is similar to the \untilP operator, but allows for counterparts to exist indefinitely without ever reaching a point where $\phi_2$ holds, provided that $\phi_1$ also keeps holding indefinitely.

The dual operators are syntactically expressed by $\false := \neg \true$, $\phi_1 \land \phi_2 := \neg (\neg \phi_1 \lor \neg \phi_2)$,
and $\forall_N x.\phi := \neg \exists_N x.\neg \phi$ and similarly for edges. {\reviewcolor Note that, differently from classical \LTL, the until and the weak until operators
are not self-dual: this fact will be discussed and made explicit in~\autoref{sec:non-expressibility-first}.}

\subsection{Satisfiability}
%
%
To present the notion of satisfiability of a formula with respect to a counterpart model we introduce the definition of \emph{assignment} for a context in a world.

\begin{definition}[Assignment]\label{def:assignment}
    An \textbf{assignment} in the world $\w\in W$ for the typed context $\G$ is a pair of functions $\mu := \abr{\mu_N,\mu_E}$ such that $\mu_N : \G_N \to D(\w)_N$ and $\mu_E : \G_E \to D(\w)_E$. We use the notation $\A^\G_\w$ to indicate the set of assignments $\mu$ defined in $\w$ for the typed context $\G$.
\end{definition}

Moreover, we denote by $\mu[x \mapsto_\tau n] \in \A^{\G,(x : \tau)}_\omega$ the assignment obtained by extending the domain of $\mu$ with $n \in D(\w)_\tau$ at the variable $x \not \in \G$, omitting the type $\tau \in \{\textsf N,\textsf E\}$ whenever clear from context. We indicate with $\G,(x : \tau)$ the context $\Gamma$ extended with an additional variable $x$ with sort $\tau$.

{\reviewcolor An assignment $\mu$ for a typed context $\G$ provides the interpretation of terms-in-context whose context is (contained in) $\G$:
it allows for evaluating the free variables of the term, thus defining its semantics (with respect to that assignment).}
This will be used in \autoref{def:qltl} to provide a meaning for the equalities $n_1 =_N n_2$  and $e_1 =_E e_2$
for nodes and edges, respectively.

\begin{definition}[Assignment on terms]\label{def:assignment_terms}
Given an assignment $\mu \in \A^\G_\omega$ ,we indicate with $\mu^* := \abr{\mu^*_N,\mu^*_E}$
the interpretation of $\mu$ on a term-in-context $[\G]\ n : \textsf{N}$ or $[\G]\ e : \textsf{E}$, given inductively by the rules: $\mu^*_E(x) := \mu_E(x)$, $\mu^*_N(y) := \mu_N(y)$, $\mu^*_N(s(e)) := D(\omega)_s(\mu^*_E(e))$, and $\mu^*_N(t(e)) := D(\omega)_t(\mu^*_E(e)).$
\end{definition}

We now look at how to transport assignments over counterpart relations. The intuition is that we have to connect the nodes and the edges in the image of two assignments when there is a counterpart relation among the worlds, and the items of the underlying graphs are related point-wise.

\begin{definition}[Counterpart relations on assignments]\label{def:counterpart_on_assignment}
    Given a counterpart relation $C \in \textsf{GraphRel}(D(\w_1), D(\w_2))$ and two assignments $\mu_1 \in \A^\G_{\omega_1}$ and $\mu_2 \in \A^\G_{\omega_2}$ on the context $\G$, we say that the assignments $\mu_1$ and $\mu_2$ are \textbf{counterpart related} if $\abr{{\mu_1}_N(x), {\mu_2}_N(x)} \in C_N$ for any $x \in \G_N$ and $\abr{{\mu_1}_E(x), {\mu_2}_E(x)} \in C_E$ for any  $x \in \G_E$.
    We indicate this with the notation $\abr{\mu_1, \mu_2} \in C$.
\end{definition}

We can now introduce the notion of satisfiability of a \QLTL formula with respect to a trace $\sigma$ and an assignment $\mu$.

\begin{definition}[\QLTL satisfiability]\label{def:qltl}
    Given a \QLTL formula-in-context $[\G]\phi$, a trace $\sigma = (C_0,C_1,\dots)$, and an assignment $\mu \in \A^\G_{\omega_0}$ in the first world of $\sigma$, we inductively define the \emph{satisfiability relation} as follows
    \begin{itemize}
        \setlength\itemsep{0.2em}
        \item $\sigma, \mu \vDash \true$;
        \item $\sigma, \mu \vDash e_1 =_E e_2$ if $\mu^*_E(e_1) = \mu^*_E(e_2)$;
        \item $\sigma, \mu \vDash n_1 =_N n_2$ if $\mu^*_N(n_1) = \mu^*_N(n_2)$;
        \item $\sigma, \mu \vDash \neg \phi$ if $\sigma, \mu \not \vDash \phi$;
        \item $\sigma, \mu \vDash \phi_1 \lor \phi_2$ if $\sigma, \mu \vDash \phi_1$ or $\sigma, \mu \vDash \phi_2$;
        \item $\sigma, \mu \vDash \exists_N x.\phi$ if there is a node $n \in D(\omega_0)_N$ such that $\sigma, \mu[x \mapsto n] \vDash \phi$;
        \item $\sigma, \mu \vDash \exists_E x.\phi$ if there is an edge $e \in D(\omega_0)_E$ such that $\sigma, \mu[x \mapsto e] \vDash \phi$;
        \item $\sigma, \mu \vDash \next \phi$ if there is $\mu_1 \in \A^\G_{\omega_1}$ such that $\abr{\mu,\mu_1} \in C_0$ and $\sigma_1, \mu_1 \vDash \phi$;
        \item $\sigma, \mu \vDash \phi_1 \until \phi_2$ if there is an $\bar n \geq 0$ such that
              \begin{enumerate}
                  \item for any $i < \bar n$, there is $\mu_i \in \A^\G_{\w_i}$ such that $\abr{\mu, \mu_i} \in  C_{\leq i}$ and $\sigma_i, \mu_i \vDash \phi_1$;
                  \item there is $\mu_{\bar n} \in \A^\G_{\w_{\bar n}}$ such that $\abr{\mu,\mu_{\bar n}} \in C_{\leq \bar n}$ and $\sigma_{\bar n}, \mu_{\bar n} \vDash \phi_2$;
              \end{enumerate}
        \item $\sigma, \mu \vDash \phi_1 \wuntil \phi_2$ if one of the following holds
              \begin{itemize}
                  \item the same conditions for $\phi_1 \until \phi_2$ apply; or
                  \item for any $i$ there is $\mu_i \in \A^\G_{\w_i}$ such that $\abr{\mu, \mu_i} \in  C_{\leq i}$ and $\sigma_i, \mu_i \vDash \phi_1$.
              \end{itemize}
    \end{itemize}
\end{definition}

\subsection{Examples}
\label{ex:example_algebraic_graphs}

We provide some examples of satisfiability for \QLTL formulae on the running example in \autoref{fig:counterpart_w_model_example} to illustrate how our counterpart semantics works in practice.
Take for example the trace given by $\sigma := (C_0,C_1,C_3,C_3,\dots)$, thus considering the case where $e_4$ is the only edge preserved when transitioning from $\omega_1$.

\begin{example}[Allocation and deallocation]  As anticipated in \autoref{sec:counterpart_semantics}, one of the main advantages of a counterpart semantics is the possibility to reason about existence, deallocation, duplication, and merging of elements in the system and its evolution. Consider for example the following shorthand formulae

    \[\begin{array}{rcl}
        \textbf{present}_\tau(x) & := & \exists_\tau y. x =_\tau y,\\
        \textbf{nextPreserved}_\tau(x) & := & \textbf{present}_\tau(x) \land \next \textbf{present}_\tau(x), \\
        \textbf{nextDealloc}_\tau(x) & := & \textbf{present}_\tau(x) \land \neg \next \textbf{present}_\tau(x). \\
    \end{array}\]

The formula $\textbf{present}_\tau(x)$ captures the existence of a node or an edge at the current moment. We can combine this predicate with the \emph{next} operator to talk about elements that are present in the current world and that will still be present at the next step, which we condense with the $\textbf{nextPreserved}_\tau(x)$ formula. We can similarly refer to elements that are now present but that will be deallocated at the next step by considering the formula $\textbf{nextDealloc}_\tau(x)$. Indeed we have

\vspace{-1em}
\[\begin{array}{l@{\hspace{0.06em}}l}
    \sigma_0, \{x : \textsf E \mapsto e_0 & \} \vDash \textbf{nextPreserved}_{E}(x); \\
    \sigma_0, \{x : \textsf N \mapsto n_1 & \} \vDash \textbf{nextPreserved}_{N}(x); \\
    \sigma_0, \{x : \textsf E \mapsto e_2 & \} \not \vDash \textbf{nextPreserved}_{E}(x); \\
\end{array}\begin{array}{l@{\hspace{0.06em}}l}
    \sigma_1, \{x : \textsf N \mapsto n_3 & \} \not \vDash \textbf{nextDealloc}_{N}(x)\\
    \sigma_1, \{x : \textsf E \mapsto e_3 & \} \vDash \textbf{nextDealloc}_{E}(x)\\
    \sigma_1, \{x : \textsf E \mapsto e_4 & \} \not \vDash \textbf{nextDealloc}_{E}(x)\\
\end{array}\]

\end{example}

\begin{example}[Graph structure] Moreover, our syntax allows us define formulae that exploit the algebraic structure of our graphs, and combine them with the temporal operators to state properties about how the graph evolves in time. We illustrate this by providing the following formulae
    \[\begin{array}{rcl}
        \textbf{loop}(e) & := & s(e) =_N t(e),\\
        \textbf{hasLoop}(n) & := & \exists_{E} e.s(e) =_{N} n \land \textbf{loop}(e),\\
        \textbf{composable}(x,y) & := & t(x) =_N s(y)\\
        \textbf{haveComposition}(x,y) & := & \textbf{composable}(x,y) \land \exists_E e.(s(e) =_N s(x) \land t(e) =_N t(y))\\
        \textbf{adjacent}(x,y) & := & \exists_E e. ((s(e) =_N x \land t(e) =_N y) \lor (t(e) =_N x \land s(e) =_N y))\\
    \end{array}\]
    which capture, respectively, the following scenarios: we can check whether a given edge of the graph is a loop with $\textbf{loop}(x)$, or verify with $\textbf{hasLoop}(x)$ that the only node having a loop is $n_5$; alternatively, we can express this fact by stating that the loop belongs to the entire world using $\textbf{hasLoop}$
    \[\begin{array}{l@{\hspace{0.02em}}l}
        \sigma_0, \{x \mapsto e_0 & \} \not \vDash \textbf{loop}(x);\\
        \sigma_1, \{x \mapsto e_3 & \} \not \vDash \textbf{loop}(x);\\
        \sigma_2, \{x \mapsto e_5 & \}      \vDash \textbf{loop}(x);\\
    \end{array}\begin{array}{l@{\hspace{0.02em}}l}
        \sigma_0, \{x \mapsto n_0 & \} \not \vDash \textbf{hasLoop}(x);\\
        \sigma_1, \{x \mapsto n_3 & \} \not \vDash \textbf{hasLoop}(x);\\
        \sigma_2, \{x \mapsto n_5 & \}      \vDash \textbf{hasLoop}(x);\\
    \end{array}\begin{array}{l@{\hspace{0.02em}}l}
        \sigma_0, \{ & \} \not \vDash \exists_N x. \textbf{hasLoop}(x)\\
        \sigma_1, \{ & \} \not \vDash \exists_N x. \textbf{hasLoop}(x)\\
        \sigma_2, \{ & \}      \vDash \exists_N x. \textbf{hasLoop}(x)\\
    \end{array}\]
    Since $\exists_N x. \textbf{hasLoop}(x)$ is a closed formula, the empty assignment is the only one the formula can be valued on.
    Thus, we obtain the classical notion of a formula simply providing the binary information of being true or false, with no choice of individuals needed to satisfy it.
    Finally, we can express some properties about the existence of intermediate nodes and composability of edges in the graph
    \[\begin{array}{l@{\hspace{0.02em}}l}
        \sigma_0, \{x \mapsto n_0, y \mapsto n_1 & \} \vDash \textbf{adjacent}(x,y);\\
        \sigma_1, \{x \mapsto n_3, y \mapsto n_4 & \} \vDash \textbf{adjacent}(x,y);\\
        \sigma_1, \{x \mapsto n_3, y \mapsto n_4 & \} \vDash \next \textbf{adjacent}(x,y);\\
    \end{array}\begin{array}{l@{\hspace{0.02em}}l}
        \sigma_0, \{x \mapsto e_0, y \mapsto e_1 & \} \vDash \textbf{composable}(x,y);\\
        \sigma_0, \{x \mapsto e_0, y \mapsto e_1 & \} \vDash \next \textbf{composable}(x,y);\\
        \sigma_1, \{x \mapsto e_3, y \mapsto e_4 & \} \not \vDash \next \textbf{composable}(x,y);\\
    \end{array}\]\[\begin{array}{l@{\hspace{0.02em}}l}
        \sigma_0, \{x \mapsto e_0, y \mapsto e_1 & \} \not \vDash \textbf{haveComposition}(x,y);\\
        \sigma_0, \{x \mapsto e_0, y \mapsto e_1 & \} \not \vDash \next \textbf{haveComposition}(x,y);\\
        \sigma_2, \{x \mapsto e_5, y \mapsto e_5 & \} \vDash \next \textbf{haveComposition}(x,y);\\
    \end{array}\]
\end{example}

\begin{remark}[\emph{Eventually} and \emph{always} operators]\label{rem:always_eventually_operator}
    As in \LTL, we can define the additional \emph{eventually} \eventuallyP and \emph{always} \alwaysP operators as $\eventually \phi := \true \until \phi$ and $\always \phi := \phi \wuntil \false$, respectively. Alternatively, their semantics can be presented directly as
    \begin{itemize}
       \item $\sigma, \mu \vDash \eventually \phi$ if  there are $i \geq 0$ and $\mu_i \in \A^\G_{\w_i}$ s.t. $\abr{\mu, \mu_i} \in  C_{\leq i}$ and $\sigma_i, \mu_i \vDash \phi$.
       \item $\sigma, \mu \vDash \always \phi$ if for any $i \geq 0$ there is $\mu_i \in \A^\G_{\w_i}$ s.t. $\abr{\mu, \mu_i} \in  C_{\leq i}$ and $\sigma_i, \mu_i \vDash \phi$.
   \end{itemize}
\end{remark}

\newpage
\begin{example}[Temporal evolution]
    We can use these operators to express the evolution of the graph after an unspecified amount of steps
    \[\begin{array}{rcl}
        \textbf{willMerge}_\tau(x,y) & := & x \not =_\tau y \land \eventually (x =_\tau y), \\
        \textbf{alwaysPreserved}_\tau(x) & := & \always \textbf{present}_\tau(x). \\
        \textbf{willBecomeLoop}(e) & := & \neg \textbf{loop}(e) \land \eventually\textbf{loop}(e)\\
    \end{array}\]

    In the example in \autoref{fig:counterpart_w} for the same trace $\sigma = (C_0,C_1,C_3,\dots)$ we have
    \[
        \begin{array}{l@{\hspace{0.02em}}l}
            \sigma_0, \{ & \} \vDash \exists n. \exists m. \textbf{willMerge}_N(n,m) ,\\
            \sigma_0, \{ & \} \not \vDash \forall e. \textbf{alwaysPreserved}_E(e) ,\\
            \sigma_0, \{ & \} \vDash \exists e. \textbf{willBecomeLoop}(e) ,\\
            \sigma_0, \{ & \} \vDash (\exists e.s(e) \not = t(e)) \until (\exists x.\forall y. x = y),\\
        \end{array}\begin{array}{l@{\hspace{0.02em}}l}
            \sigma_0, \{ & \} \not \vDash \forall e. \eventually \textbf{loop}(e) ,\\
            \sigma_0, \{ & \} \vDash \exists e. \eventually \always \textbf{loop}(e),\\
            \sigma_0, \{ & \} \vDash \exists x. \exists y. \neg \eventually \textbf{composable}(x,y) ,\\
            \sigma_0, \{ & \} \vDash (\exists e.s(e) = t(e)) \wuntil \neg (\exists x.\textbf{loop}(e)).\\
        \end{array}
    \]

\end{example}

\begin{remark}[Quantifier elision for unbound variables]\label{rem:quantifier_elision}
    A relevant difference with standard quantified logics is that in \QLTL we cannot elide quantifications where the variable introduced does not appear in the subformula.
    {\reviewcolor Assuming  $\equiv$ to denote semantical equivalence and} taking any $\phi$ with $x \not \in \fv(\phi)$, we have that in general $\exists x.\phi \not \equiv \phi$ and,
    similarly, $\forall x.\phi \not \equiv \phi$.
    More precisely, the above equivalences hold whenever $\phi$ does not contain any temporal operator and the current world $D(\omega)$ being considered is not empty.

    Consider a world $\w$ with a single node $D(\w)_N = \set{s}$, no edges, and a single looping counterpart relation
    $\C \abr{\w,\w} = \set{C}$ where $C = \emptyset$ is the empty counterpart relation. The trace is given by $\sigma = (C,C,\dots)$. By taking the empty assignment $\emptymu$ and the closed formula $\phi = \next(\true)$, one can easily check that $\sigma, \emptymu \vDash \next(\true)$, but $\sigma, \emptymu \not \vDash \exists_N x.\next(\true)$.
    The reason is that, once an assignment is extended with some element, stepping from one world to the next one requires every individual of the assignment to be preserved and have a counterpart in the next world.
    Alternatively, we could have restricted assignments in the semantics so that counterparts are required only for the free variables occurring in the formula.
    For example, the definition for the \emph{next} \nexteP operator would become
    \begin{itemize}
        \item $\sigma, \mu \vDash \next \phi$ if there is $\mu_1 \in A^{\fv(\phi)}_{\w_1}$ such that $\abr{\mu_{\mid \fv(\phi)}, \mu_1} \in C_0$ and $\sigma_1, \mu_1 \vDash \phi$
    \end{itemize}
    For ease of presentation in this work and with respect to our Agda implementation, we consider the case where all elements in the context have a counterpart. 
\end{remark}

\begin{remark}[\emph{Until} and \emph{weak until} are incompatible]\label{sec:non-expressibility-first}
    In standard \LTL, the \emph{until}~\untilP and \emph{weak until}~\wuntilP operators have the same expressivity, and can be defined in terms of each other by the equivalences $\phi_1 \until \phi_2 \equiv_\LTL \neg (\neg \phi_2 \wuntil (\neg \phi_1 \land \neg \phi_2)), \phi_1 \wuntil \phi_2 \equiv_\LTL \neg(\neg \phi_2 \until (\neg \phi_1 \land \neg \phi_2))$.
    However, this is not the case in \QLTL. Similarly, it might at first seem reasonable to define the standard \emph{always} operator in \QLTL with $\always \phi := \neg \eventually \neg \phi$. However, this definition does not align with the semantics provided in \autoref{rem:always_eventually_operator}. This characteristic of \QLTL is again due to the fact that we are in the setting of (possibly deallocating) relations, and we formally explain and present an intuition for this when we introduce the PNF semantics in \autoref{sec:pnf}. The \LTL equivalences can be obtained again by restricting to models whose counterpart relations are total functions: this allows us to consider a unique trace of counterparts that are always defined, which brings our models back to a standard \LTL-like notion of trace. 
\end{remark}

\section{Positive normal form for \QLTL}
\label{sec:pnf}

Positive normal forms are a standard presentation of temporal logics: they can be used to simplify constructions and algorithms on both the theoretical and implemention side~\cite{regular-vacuity:2005,tableau-finite-ltl:2022}. Their use is crucial for semantics based on fixpoints, such as in \cite{counterpart-semantics-second-order-mu-calculus:2012}, while still preserving the expressiveness of the original presentation. As we remark in \autoref{ch:agda_formalisation}, providing a negation-free semantics for our logic also ensures that it can be more easily manipulated in a proof assistant where definitions and proofs are \emph{constructive}. Moreover, the positive normal form conversion serves as a concrete procedure that can be used interactively to automatically convert formulae into their positive normal form, which is proven in the proof assistant to be correct. In this section we present an explicit semantics for the positive normal form of \QLTL, which we denote as \PNF.

\subsection{Semantics of \PNF}
\label{sec:pnC_semantics}

As observed in 
\autoref{sec:non-expressibility-first}, to present the positive normal form we need additional operators to adequately capture the negation of temporal operators. Thus, we introduce a new flavour of the next operator, called \emph{next-forall} \nextfP. Similarly, we have to introduce a dual for \emph{until}~\untilP and \emph{weak until}~\wuntilP , which we indicate as the \emph{then}~\thenP and \emph{until-forall}~\funtilP operators, respectively.

\begin{definition}[\QLTL in PNF]\label{def:pnC_syntax}
    Let $\Gamma$ be a typed context on the set of variables $\X$.
    The set $\F^\PNF_\Gamma$ of formulae of \QLTL in \textbf{positive normal form} is generated by the following rules
    \[ \psi := \true
    \mid e_1 =_E e_2
    \mid n_1 =_N n_2 \]
    \[
        \phi := \psi
        \mid \neg \psi
        \mid \phi \lor \phi
        \mid \phi \land \phi
        \mid \exists_\tau x. \phi
        \mid \forall_\tau x. \phi \\
        \mid \nexte \phi
        \mid \nextf \phi
        \mid \phi \until \phi
        \mid \phi \funtil \phi
        \mid \phi \wuntil \phi
        \mid \phi \then \phi.
    \]
    where $[\G]\ e_i : \textsf{E}$ and $[\G]\ n_i : \textsf{N}$ for $i = 1, 2$ and $\tau \in \{\textsf{N},\textsf{E}\}$.
\end{definition}

The intuition for the \emph{next-forall} \nextfP operator is that it allows us to capture the case where a counterpart of an individual does not exist at the next step: if any counterpart exists, it is required to satisfy the formula $\phi$.

Similarly to the \emph{until}~\untilP operator, the \emph{until-forall}~\funtilP operator allows us to take a sequence of graphs where $\phi_1$ is satisfied for some steps until $\phi_2$ holds. The crucial observation is that \emph{every} intermediate counterpart satisfying $\phi_1$ and the conclusive counterparts must satisfy $\phi_2$. Such counterparts are not required to exist, and indeed any trace consisting of all empty counterpart relations always satisfies both \funtilP and \thenP. Similarly to the \emph{weak until}~\wuntilP operator, the \emph{then}~\thenP operator corresponds to a \emph{weak until-forall} and can be validated by a trace where all counterparts satisfy $\phi_1$ without ever satisfying $\phi_2$. 

We now provide a satisfiability relation for \PNF formulae by specifying the semantics just for the additional operators, omitting the ones that do not change.

\begin{definition}[\QLTL in PNF satisfiability]\label{def:pnf}
    Given a \QLTL formula-in-context $[\G]\phi$ in positive normal form, a trace $\sigma = (C_0,C_1,\dots)$, and an assignment $\mu \in \A^\G_{\omega_0}$ in the first world of $\sigma$, we inductively define the \emph{satisfiability relation} with respect to the additional operators as follows
    \begin{itemize}
        \item $\sigma, \mu \vDash \neg \psi$ if $\sigma, \mu \not \vDash \psi$;
        \item $\sigma, \mu \vDash \phi_1 \land \phi_2$ if $\sigma, \mu \vDash \phi_1$ and $\sigma, \mu \vDash \phi_2$;
        \item $\sigma, \mu \vDash \forall_\tau x.\phi$ if for any $s \in D(\omega_0)_\tau$ we have that $\sigma, \mu[x \mapsto s] \vDash \phi$;
        \item $\sigma, \mu \vDash \nextf \phi$ if for any $\mu_1 \in \A^\G_{\w_1}$ such that $\abr{\mu, \mu_1} \in C_0$ we have that $\sigma_1, \mu_1 \vDash \phi$;
        \item $\sigma, \mu \vDash \phi_1 \funtil \phi_2$ if there is an $\bar n \geq 0$ such that
              \begin{enumerate}
                  \item for any $i < \bar n$ and $\mu_i \in \A^\G_{\w_i}$ such that $\abr{\mu, \mu_i} \in  C_{\leq i}$ we have $\sigma_i, \mu_i \vDash \phi_1$;
                  \item for any $\mu_{\bar n} \in \A^\G_{\w_{\bar n}}$ such that $\abr{\mu,\mu_{\bar n}} \in C_{\leq \bar n}$ we have  $\sigma_{\bar n}, \mu_{\bar n} \vDash \phi_2$;
              \end{enumerate}
        \item $\sigma, \mu \vDash \phi_1 \then \phi_2$ if one of the following holds
              \begin{itemize}
                  \item the same conditions for $\phi_1 \funtil \phi_2$ apply; or
                  \item for any $i$ and $\mu_i \in \A^\G_{\w_i}$ such that $\abr{\mu, \mu_i} \in  C_{\leq i}$ we have $\sigma_i, \mu_i \vDash \phi_1$.
              \end{itemize}
    \end{itemize}
\end{definition}


\begin{figure}[H]
    \centering
    \vspace{-1.3em}
    \input{running_example_graph2.tikz}
    \vspace{-2.3em}
    \caption{Graphical representation of a counterpart model.}
    \label{fig:counterpart_w_model_example2}
\end{figure}
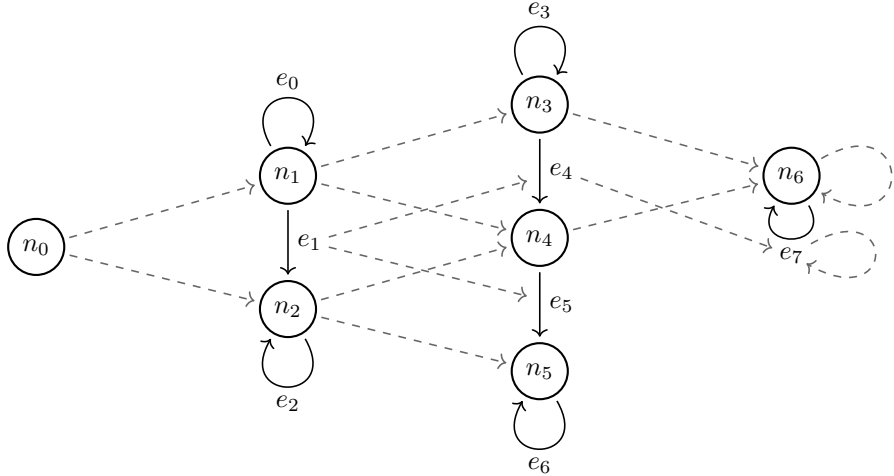

\begin{example}
We illustrate with the example in \autoref{fig:counterpart_w_model_example2} the possibility for a counterpart relation to duplicate both edges and nodes of a graph, as well as providing some concrete cases for the new operators defined in \autoref{def:pnf}. For example, we have that $\sigma_0, \{x \mapsto n_0 \} \vDash \textsf{A}(\textbf{hasLoop}(x))$, but $\sigma_1, \{x \mapsto n_1 \} \not \vDash \textsf{A}(\textbf{hasLoop}(x))$ since $n_4$ is a counterpart of $n_1$ but does not have a loop. Moreover, $\sigma_1, \{x \mapsto e_1 \} \vDash \textbf{hasLoop}(s(x))\textsf{U}(\textbf{loop}(x))$, but $\sigma_1, \{x \mapsto e_1 \} \not \vDash \textbf{hasLoop}(s(x))\textsf{F}(\textbf{loop}(x))$ because we also require for $e_5$ to have a loop at its source since it is a counterpart of $e_1$. Notice that $\sigma_2, \{x \mapsto e_5 \} \vDash \textsf{A}(\textbf{loop}(x))$ since there is no counterpart at the next step, and indeed we similarly have that $\sigma_2, \{x \mapsto n_5 \} \vDash \textbf{hasLoop}(x)\textsf{F}(\textsf{false})$. Finally, we have that $\sigma_2, \{x \mapsto e_4 \} \vDash \textbf{hasLoop}(s(x))\textsf{T}(\neg \textbf{loop}(x))$ because the intermediate condition always holds.

\end{example}

\subsection{Negation of \QLTL and \PNF}\label{ex:negation_both}

The crucial observation that validates the PNF presented in \autoref{sec:pnf} is that the negation of \emph{next} $\nexte\phi$, \emph{until}~\untilP, and  \emph{weak until}~\wuntilP formulae can now be expressed inside the logic. We indicate with $\vDash_{\QLTL}$ and $\vDash_\PNF$ the satisfiability relations for formulae in standard \QLTL and \QLTL in PNF, respectively.

\begin{proposizia}[Negation is expressible in PNF]\label{pro:neg_expressible}
    \agda{https://github.com/iwilare/algebraic-temporal-logics/blob/main/PNF/Relational/Negation.agda}{Relational.Negation}\\
    Let $\psi, \psi_1, \psi_2$ be atomic formulae in PNF. Then we have
    \[\begin{array}{llll}
        \forall \sigma, \mu \in \A^\G_{\omega_0}.&\sigma, \mu \vDash_{\QLTL} \neg \nexte( \psi) & \iff & \sigma, \mu \vDash_\PNF \nextf( \neg \psi) \\
        \forall \sigma, \mu \in \A^\G_{\omega_0}.&\sigma, \mu \vDash_{\QLTL} \neg (\psi_1 \until \psi_2) & \iff & \sigma, \mu \vDash_\PNF (\neg \psi_2) \then (\neg \psi_1 \land \neg \psi_2)\\
        \forall \sigma, \mu \in \A^\G_{\omega_0}.&\sigma, \mu \vDash_{\QLTL} \neg (\psi_1 \wuntil \psi_2) & \iff & \sigma, \mu \vDash_\PNF (\neg \psi_2) \funtil (\neg \psi_1 \land \neg \psi_2).\\
    \end{array}\]
\end{proposizia}

A converse statement that similarly expresses the negation of these new operators in \PNF does not hold: the only exception is the easy case of the \emph{next-forall} \nextfP operator, whose negation directly corresponds with the \emph{next} \nextP operator.

\newcommand{\notiff}{%
  \mathrel{{\ooalign{\hidewidth$\not\phantom{"}$\hidewidth\cr$\iff$}}}}

\begin{proposizia}[Negation of new operators is \emph{not} in PNF]\label{pro:negation_new_operators}
    Let $\psi, \psi_1, \psi_2$ be atomic formulae in PNF. Then we have
    \[\begin{array}{llll}
    \forall \sigma, \mu \in \A^\G_{\omega_0}.&\sigma, \mu \not \vDash_\PNF \nextf( \psi) & \iff & \sigma, \mu \vDash_\PNF \nexte( \neg \psi)\\
    \forall \sigma, \mu \in \A^\G_{\omega_0}.&\sigma, \mu \not \vDash_\PNF \psi_1 \then \psi_2 & \notiff & \sigma, \mu \vDash_\PNF (\neg \psi_2) \until (\neg \psi_1 \land \neg \psi_2)\\
    \forall \sigma, \mu \in \A^\G_{\omega_0}.&\sigma, \mu \not \vDash_\PNF \psi_1 \funtil \psi_2 & \notiff & \sigma, \mu \vDash_\PNF (\neg \psi_2) \wuntil (\neg \psi_1 \land \neg \psi_2).\\
    \end{array}\]
\end{proposizia}
\begin{proof}
See \autoref{appendix:negation_new_operators}.
\end{proof}

Let us now go back to \autoref{pro:neg_expressible}. We exploit the correspondence between operators given there to define a translation
$\overline{\mathmakebox[1em]{\cdot}} : \F^\QLTL \to \F^\PNF$ from the \QLTL syntax presented in \autoref{def:qltl} to the current one in PNF.
What is noteworthy is that such translation preserves the
equivalence of formulae. 

\begin{theorem}[PNF equivalence]\label{thm:pnC_theorem}
    \agda{https://github.com/iwilare/algebraic-temporal-logics/blob/main/PNF/Relational/Conversion.agda}{Relational.Conversion}
    Let $\overline{\mathmakebox[1em]{\cdot}} : \F^\QLTL \to \F^\PNF$ be the syntactical translation that replaces negated temporal operators
    with their equivalent ones in PNF by pushing negation down to atomic formulae. For any \QLTL formula $[\G]\phi \in \F^\QLTL$ we have
    \[\begin{array}{llll}
        \forall \sigma, \mu \in \A^\G_{\omega_0}.&\sigma, \mu \vDash_{\QLTL} \phi & \iff & \sigma, \mu \vDash_\PNF \overline{\mathmakebox[1em]{\phi}}.
    \end{array}\]
\end{theorem}

Contrary to what happens in \LTL, the usual expansion laws where each operator is defined in terms of itself do not hold in \QLTL for the case of counterpart relations, as shown by the following result.

\begin{proposizia}[Expansion laws do \emph{not} hold in \QLTL]\label{pro:exp_law_not}
We have the following inequalities in \PNF
    \[%
        \arraycolsep=10pt
        \begin{array}{ll}
            \phi_1 \until \phi_2 \not \equiv \phi_2 \lor (\phi_1 \land \nexte(\phi_1 \until \phi_2))   &
            \phi_1 \funtil \phi_2 \not \equiv \phi_2 \lor (\phi_1 \land \nextf(\phi_1 \funtil \phi_2))  \\
            \phi_1 \wuntil \phi_2 \not \equiv \phi_2 \lor (\phi_1 \land \nexte(\phi_1 \wuntil \phi_2)) &
            \phi_1 \then \phi_2 \not \equiv \phi_2 \lor (\phi_1 \land \nextf(\phi_1 \then \phi_2)).       \\
        \end{array}
    \]%
\end{proposizia}
\begin{proof}
    See \autoref{appendix:expansion_laws}.
\end{proof}

\begin{remark}[Functional counterpart relations]
The previous results can be reframed in the case in which each counterpart relation is a \emph{partial function}, following the definition of counterpart models given in \cite{presheaf-semantics-qtl:2021,counterpart-semantics-second-order-mu-calculus:2012}. It turns out that under the assumption of partial functions we recover all the equivalences stated in \autoref{pro:negation_new_operators} \agda{https://github.com/iwilare/algebraic-temporal-logics/blob/main/PNF/Functional/Negation.agda}{Functional.Negation} as well as the expansion laws of \autoref{pro:exp_law_not} \agda{https://github.com/iwilare/algebraic-temporal-logics/blob/main/PNF/Functional/ExpansionLaws.agda}{Functional.ExpansionLaws}. The latter can be used to provide a presentation of the temporal operators as \emph{least fixpoint} (\untilP, \funtilP) and \emph{greatest fixpoints} (\wuntilP, \thenP) of a suitably defined operator; this definition based on fixpoints would coincide with the semantics of the operators given in \autoref{def:pnf} only in the case of partial (or total) functions.
\end{remark}

\section{Agda formalisation}
\label{ch:agda_formalisation}

This section presents an overview of an additional contribution of this work: a complete formalisation of the semantics of \QLTL and its \PNF using the dependently typed programming language and proof assistant Agda \cite{dependently-typed-programming-agda:2009}. We provide a brief exposition and usage of our development in {\textcolor{red}{Appendix C}} by showing how the temporal evolution of the running example in \autoref{fig:counterpart_w_model_example} can be concretely modelled in Agda.
The complete formalisation of the logic along with the PNF results is available at \url{https://github.com/iwilare/algebraic-temporal-logics}.

\subsection{Formalisation aspects}\label{sec:formalisation_aspects}
{\reviewcolor
Our formalisation work consists in the mechanisation of all the aspects presented in the paper: we start by defining the notion of counterpart relations and traces of relational morphisms as models of the logic, and provide a representation for (well-typed and well-scoped) syntax for formulae of \QLTL and \PNF along with their satisfiability semantics. Then, we provide a conversion function from \QLTL to \PNF along with proofs of correctness and completeness of the procedure; finally, using the defined framework, we prove among other equivalences the relevant expansion laws introduced in \autoref{ex:negation_both} for the functional setting.

For the sake of presentation in this paper we restricted our attention to graph signatures, and the modelling example presented in {\textcolor{red}{Appendix C}} is
indeed instantiated on the signature of directed graphs.  However, we remark that our implementation is general enough to model
algebras over any generic multi-sorted signature. In particular, this means that, by specifying a suitably defined signature, the class of models (and formulae) considered by the logic can be extended to the case of any graphical formalism
 that admits an algebraic representation on a multi-sorted signature, such as multigraphs, non-directed graphs, and the typed attributed graphs of~\cite{metric-temporal-graph-logic-TAG:2019}.

Moreover, given the constructive interpretation of the formalisation, proving that the correctness and completeness of \PNF with respect to \QLTL also doubles-down as concrete procedure that can convert formulae into their positive normal form version, while at the same time providing a proof of the correctness of the conversion. We describe now how the main components provided by our formalisation can be employed by the user to interact with the proof assistant.

\vspace{-0.75em}
    \paragraph{Signature definition.} Using the definitions given in our formalisation, the user can write their own algebraic signature that will be used to represent the system of interest as algebras on the signature. For example, by defining the signature of graphs $\textsf{Gr}$ the user can reason on the temporal evolution of graphs, using (relational) graph homomorphisms as counterpart relations between worlds.
\vspace{-0.75em}
    \paragraph{Formula construction.} After having provided the signature of interest, the user can construct formulae using the full expressiveness of \QLTL and can reason on equality of terms constructing according to the signature. This allows the user to express properties that combine both logical quantifiers as well as exploiting the specific structure of the system, possibly composing and reusing previously defined formulae. The infrastructure provided by the formalisation is such that the formulae constructed by the user are inherently checked to be well-scoped and well-typed with respect to the sorts of the signature, e.g. edges and nodes in the case of graphs. The user can freely use negation in formulae, and can (optionally) use the procedures we defined to automatically convert formulae to their \PNF, which we have seen in \autoref{sec:pnf} how can be particularly counterintuitive in the counterpart setting with respect to standard temporal logics.
\vspace{-0.75em}
    \paragraph{Model definition.} The models of the system at various time instances can be constructed by the user, following again the signature provided. Then, the user specifies a series of symbolic worlds and indicates the possible transitions that can be taken by defining a relation on the worlds. Then, an algebra of the signature must be assigned to each world, and the connection of worlds is translated into a morphism between the algebras which the user provides. The transitions of the models are checked by Agda to preserve the algebraic structure of the worlds considered, thus corresponding to the notion of graph morphisms; this step is relatively straightforward as the automation available in Agda helps with proving the structure-preservation of the maps. Traces between worlds are defined using a coinductive definition of traces using sized types~\cite{sized-types}, thus allowing for infinite (repeating) traces to be modelled and defined by the user.
\vspace{-0.75em}
    \paragraph{Validation of formulae in the model.} Using the library the user can prove that a specific model satisfies a given formula; our formalisation automatically simplifies the goal that must be proven to verify the formula, and the user is guided by the proof assistant by automatically constructing the skeleton of the proof term.
}

\subsection{Intuitionistic proof assistant}

In our setting, some crucial usability issues need to be mentioned. Agda is a proof assistant based on the \emph{intuitionistic} interpretation of mathematics~\cite{proof-and-types:1989}. 
This means that some useful logical principles often used in the setting of temporal logics are \emph{not provable} in the system, such as the law of excluded middle or the De Morgan laws to switch connectives and quantifiers whenever negation appears in subformulae. 
Thus, without assuming these logical principles, the embedding of our temporal logic \QLTL would actually be restricted to the \emph{intuitionistic} fragment; in practice, this is not particularly problematic since classical reasoning can simply be assumed as axiom, and allows the equivalences mentioned  to be recovered. This, however, would be undesirable from the user's perspective, as they would have to explicitly use these classical axioms in their proofs.
In order to tackle these usability aspects and the treatment of negation in the intuitionistic setting, we take the following approach: the formulae of the logic are expressed in Agda using a full \emph{positive normal form} similar to the one presented in \autoref{sec:pnf}, which we have proven correct in Agda in \autoref{thm:pnC_theorem} by postulating classical principles. This effectively shifts the burden of dealing (classically) with negation from the user to the implementer, while also giving them complete accessibility over the extended set of correct quantifiers. Moreover, the correctness proof of the conversion to \PNF constitutes both a theoretical guarantee that no expressive power is either gained or lost in the presentation of the logic, as well as a concrete algorithm that the user can execute to convert their formulae into equivalent ones in \PNF, for which validity can be easier to prove.

\subsection{Automation}

Embedding a temporal logic in a proof assistant allows the user to exploit the \emph{assistant} aspect of the tool, for example, by aiding the user in showing (or even prove automatically) that a certain formula in a model is satisfied or not.

In Agda, this automation aspect is limited, especially if compared to proof assistants where automation and the use of tactics is a core aspect of the software environment, such as Coq \cite{coq:2016}, Lean \cite{lean:2021}, and Isabelle \cite{isabelle:2002}.
The Agda synthesizer Agsy \cite{agsy:2004} is the main helper tool in Agda implementing a form of automated proof search. Unfortunately, Agsy only provides general-purpose searching procedures and its theorem proving capabilities are nowhere near those of specialised model checking algorithms.
Still, the goal-oriented interactivity available in Agda is an invaluable tool in proving theorems step-by-step and \emph{manually} verify formulae in our setting, and the assisted introduction of constructors allows the user to quickly generate the proof structure needed to validate temporal formulae.

%

\section{Related works}
Up to the early 2010s, there has been a series of papers devoted to some variants of quantified logics for expressing properties of graphs
and of graph evolutions. Our models are inspired by the counterpart-based logics explored in the context of a $\mu$-calculus with fixpoints in
\cite{counterpart-semantics-second-order-mu-calculus:2012}, and we refer there for an overview of and a comparison with the by-then current proposals,
such as the well-known \cite{tgl-vfts}, all favouring an approach based on universal domains.
Among the follow-ups of the works mentioned there, there is~\cite{GhamarianMRZZ12} and the related~\cite{SmidR13,ZambonR18},
which further explore one of the relevant tools developed in the graph community, GROOVE.
To some extent, the present paper and its companion~\cite{presheaf-semantics-qtl:2021}, which introduces the categorical semantics of second-order \QLTL,
are summarising a previous thread of research concerning counterpart models, including its implementation. And in fact, the categorical
semantics for counterpart models appears of interest in itself in the literature on modal logics, as witnessed by the works surveyed in~\cite{presheaf-semantics-qtl:2021}.

{\reviewcolor
\vspace{0.5em}
Concerning the formalisation of temporal logics in (constructive) proof assistants, the topic has a long history, see e.g. \cite{axiomatization-ltl-in-coc:2003,verified-model-checker-modal-mu-calculus-in-coq:1998,alternating-time-temporal-logic-in-coc:2012}. A practical application and comparison with modern model checkers is in \cite{fully-verified-executable-ltl-model-checker:2014}, and a fully verified \LTL model checker is implemented in the Isabelle theorem prover. In \cite{applications-applicative-proof-search:2016}, a verified proof-search program is formalised in Agda for standard \CTL, together with a toolbox to implement well-typed proof-searching procedures; a similar embedding of constructive \LTL in Agda is provided in \cite{ltl-types-frp:2012} for the verification of functional reactive programs.
Our proof-of-concept implementation of \QLTL in Agda witnesses the possibility to move towards the formalisation of quantified temporal logics for proof assistants,
an issue sparsely tackled in the literature.

\vspace{0.5em}
Concerning graph computation models (GCMs), we  find in the literature several formalisms that use graph-specific definitions where
syntactical statements on nodes, edges, sources of edges, targets of edges, and equalities are first-class citizens in the logic to express
properties on the system under analysis.
The last decade has seen a series of papers advocating quantified temporal logics as a formalism for the specification of GCMs properties.
We offer a short review of some of the most recent proposals appeared in the literature, focussing on the dichotomy
between the universal domains and the counterpart-based approaches.
\vspace{-1.5em}
\paragraph{Graph programs/flow graphs.} The use of monadic-second order logics to prove  properties of graph-based programming languages has been advocated
in~\cite{WulandariP21,PoskittP23}, where the emphasis is placed on distilling post-conditions formulae from a  graph transformation rule and a precondition formula. A more abstract meta-model for
run-time verification is proposed in~\cite{worst-case-execution-time-calculation:2021,specification-language-consistent-model-generation:2020},
where a control flow graph can be instantiated to concrete models and the properties are
given by first-order formulae. Despite the differences, in both cases the resulting analysis is akin to the adoption of a universal domain approach.
\vspace{-0.75em}
\paragraph{Metric logics, I.} The use of traces and first-order specifications is a key ingredient of runtime verification. A relevant proposal is the use of
metric first-order temporal logic (\textsf{MFOTL})~\cite{formally-verified-monitor-MFOTL:2019,formalisation-monitoring-MFOTL:2019}, investigated with respect to
the expressiveness of suitable fragments in~\cite{real-time-policy-MFOTL:2022} or to
duality results akin to our \PNF in~\cite{relaxing-safety-MFOTL:2022}.
These logics allows to reason on the individual components of states, using (arbitrary) sets of relations as models, which allows for different kinds of graphs to be encoded.
The core difference with our line of work is that, contrary to standard models of \textsf{MFOTL}, we allow for variable domains in the temporal structure and for nodes to be created and destroyed using counterpart relations.
\vspace{-0.75em}
\paragraph{Metric logics, II.} A graph-oriented approach to \textsf{MFOTL} is given by Metric Temporal Graph Logic (\textsf{MTGL})~\cite{metric-temporal-graph-logic-TAG:2019,optimistic-pessimistic-analysis-MTGL:2020}, which allows to model properties on the structures and the attributes of the state and has been used in
the context of formal testing~\cite{formal-testing-timed-graph-transformation:2021}. Here traces are pairs of injective spans representing a rule,
 and are equivalent to our partial graph morphisms. The syntax is tailored over such rules, so that
$\phi_G$ refers to a formula over a given graph $G$, and a one step $\exists(f, \phi_{H})$ is indexed over a mono $f: G \rightarrow H$,
roughly representing the partial morphism induced by a rule. Thus, besides our use of relations, identity and preservation/deletion of elements
seem to be left implicit, and the exploration of the connection with counterpart-based \QLTL is among our future endeavours.
}

\section{Conclusions and future works}
\label{ch:conclusion}

We have seen how a set-theoretic semantics for a first-order linear-time temporal logic \QLTL can be presented in the counterpart setting. We saw how its syntax and semantics can be naturally used in an algebraic setting to express properties of direct graphs and their evolution in time, and how the notions and models presented in the previous sections can be formalised and practically experimented with in a proof assistant based on dependent type theory such as Agda. We have investigated some results on the positive normal forms of this logic in the case of relations and partial functions, and argued for their usefulness both in practice and in the case of constructive proof assistants.

\vspace{0.4em}
We identify a variety of possible expansions for our work.
\vspace{-0.7em}
\paragraph{Second-order.} Our theoretical presentation and formalisation work focuses on the first-order aspects of \QLTL. The semantics in \cite{presheaf-semantics-qtl:2021,counterpart-semantics-second-order-mu-calculus:2012} allows also for the quantification over sets of elements. This is impractical in Agda due to the typical formalisation of subsets as predicates, which would be cumbersome to present in concrete examples, e.g. when expressing universal quantification and extensional equality over subsets of elements. A possible extension could be to investigate practical encodings and possible automation techniques to introduce second-order quantification for counterpart-based temporal logics.
\vspace{-1em}
\paragraph{CTL and other logics.} The quantified temporal logics presented here focus on providing a restricted yet sufficiently powerful set of operators and structures. These logics could be extended to alternative constructs and models, such as those of \CTL \cite{temporal-and-modal-logic:1990}.
Extending our logic to more complex models seems a straightforward task, which might however cause a combinatorial explosion in the temporal operators required to obtained a positive normal form.
\vspace{-1em}
\paragraph{Automation and solvers.} We highlighted how the proofs required to validate temporal formulae need to be provided manually by the user. Considerable amount of effort has been spent in interfacing proof assistants with external solvers and checkers to both reuse existing work and algorithms and to provide more efficient alternatives to the automation given by proof assistants. The traditional way of employing proof automation is through the use of \emph{internal} and \emph{external} solvers: the first technique uses the reflection capabilities of Agda to allow a (verified) solver and proof-searching procedure to be written in Agda itself, in the spirit of \cite{applications-applicative-proof-search:2016,fully-verified-executable-ltl-model-checker:2014,auto-in-agda:2015}. The second mechanism consists in writing bindings to external programs, such as model checkers or SMT and SAT solvers, so that proving the formula or providing a counterexample is offloaded to a more efficient and specialised program. A possible extension of this work would be the implementation of either of these mechanisms to the setting of counterpart semantics. 
\vspace{-1em}
\paragraph{Finite traces.} A current trend in artificial intelligence is the study of temporal formulas over \emph{finite} traces~\cite{GiacomoV15}, due to applications in planning and reinforcement learning. Our models seem to be well-suited to tackle such a development, since each finite trace can be thought of as an infinite one terminating with a cycle in an empty graph, thus inheriting all the issues we highlighted about positive normal forms for our logic.

\bibliographystyle{splncs04}
\bibliography{bibliography}

\clearpage

\appendix
\section{Appendix}
\label{graphexample}

We provide an explicit description of the model given in \autoref{fig:counterpart_w_model_example}. The graphs $D(\omega_i)$ assigned to each world and the sets of graph morphisms $\C\abr{\omega_i,\omega_j}$ connecting them are given as

    \[\begin{array}{ll}
        D(\omega_0)_E = \set{e_0,e_1,e_2}, &
        D(\omega_0)_N = \set{n_0,n_1,n_2}; \\
        D(\omega_1)_E = \set{e_3,e_4}, &
        D(\omega_1)_N = \set{n_3,n_4}; \\
        D(\omega_2)_E = \set{e_5}, &
        D(\omega_2)_N = \set{n_5}. \\
    \end{array}\]

The sets of atomic counterpart relations connecting each pair of worlds are given as follows

    \[\begin{array}{l@{ }l}
        \C\abr{\omega_0,\omega_1} & = \set{C_0}, \\
        \C\abr{\omega_1,\omega_2} & = \set{C_1,C_2}, \\
        \C\abr{\omega_2,\omega_2} & = \set{C_1,C_2}, \\
    \end{array}\begin{array}{l@{ }l}
        \C\abr{\omega_0,\omega_0} & = \emptyset{}, \\
        \C\abr{\omega_1,\omega_1} & = \emptyset{}, \\
        \C\abr{\omega_0,\omega_2} & = \emptyset{}, \\
    \end{array}\]
    and $C_0 \in \textsf{GraphRel}(D(\omega_0),D(\omega_1))$, $C_1,C_2 \in \textsf{GraphRel}(D(\omega_1),D(\omega_2))$, $C_3 \in \textsf{GraphRel}(D(\omega_2),D(\omega_2))$ are the following graph morphisms
    \[\begin{array}{ll}
        {C_0}_N = \set{(n_4,n_0),(n_3, n_1),(n_4,n_2)}, &
        {C_0}_E = \set{(e_4, e_0),(e_3,e_1)}; \\
        {C_1}_N = \set{(n_5,n_3),(n_5,n_4)}, &
        {C_1}_E = \set{(e_5,e_4)}; \\
        {C_2}_N = \set{(n_5,n_3),(n_5,n_4)}, &
        {C_2}_E = \set{(e_5,e_3)}; \\
        {C_3}_N = \set{(n_5,n_5)}, &
        {C_3}_E = \set{(e_5,e_5)}; \\
    \end{array}\]

\section{Omitted proofs}
\label{appendixTheorems}

We include explicit proofs of previously stated theorems. We consider some very simple cases in which only nodes exist in the graph and some distinguishing property on nodes $x$ is indicated using the unary predicates $\blue(x)$ and $\red(x)$.

\subsection{Proof of \autoref{pro:negation_new_operators}}
\label{appendix:negation_new_operators}

\begin{figure}[H]
    \centering
    \input{then_counterexample.tikz}
    \caption{A counterexample model where, in the case of counterpart relations, we have that $\neg (\blue(x) \then \red(x)) \not \equiv (\neg \red(x)) \until (\neg \blue(x) \land \neg \red(x))$.}\label{fig:then_counterexample}
\end{figure}
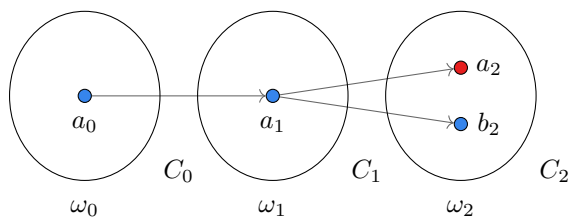

We provide a single counterexample for both the \emph{then} and \emph{until-forall} cases. Take for example the formula $\blue(x) \then \red(x)$; we consider the graphs in \autoref{fig:then_counterexample}

Clearly, we have that $\sigma, \set{x \mapsto a_0} \vDash \neg (\blue(x) \then \red(x))$. However, $\sigma, \set{x \mapsto a_0} \not \vDash (\neg \red(x)) \until (\neg \blue(x) \land \neg \red(x))$ since the \emph{until} operator requires a \emph{single} counterpart to exist where both $\neg \blue(x)$ and $\neg \red(x)$ after $n$ steps. The case of \funtilP and its negated form using \wuntilP follows similarly.

\subsection{Proof of \autoref{pro:exp_law_not}}
\label{appendix:expansion_laws}

    We provide direct counterexamples for the \emph{until} \untilP and \emph{until-forall} \funtilP cases, with the \emph{weak until} \wuntilP and \emph{then} \thenP cases obviously following.

    \begin{figure}[H]
        \centering
        \input{exp_law_counterexample_until.tikz}
        \caption{A counterexample model where, in the case of counterpart relations, we have that $\blue(x) \until \red(x) \not \equiv \red(x) \lor (\blue(x) \land \nexte(\blue(x) \until \red(x))) $.}\label{fig:exp_law_counterexample_until}
    \end{figure}

    Consider the \emph{until} case with the formula $\blue(x) \until \red(x)$. In the model shown in \autoref{fig:exp_law_counterexample_until} we have that $\sigma , \set{x \mapsto a_0} \vDash \blue(x) \until \red(x)$ since there exists a counterpart after two steps with $\red(x)$, and for all graphs before it there exists a counterpart with $\blue(x)$. However, clearly $a_0$ does not satisfy the expanded formula since neither $\sigma_1 , \set{x \mapsto a_1} \not \vDash \blue(x) \until \red(x)$ nor $\sigma_1 , \set{x \mapsto b_1} \not \vDash \blue(x) \until \red(x)$.

    \begin{figure}[H]
        \centering
        \input{exp_law_counterexample_until_forall.tikz}
        \caption{A counterexample model where, in the case of counterpart relations, we have that $\blue(x) \funtil \red(x) \not \equiv \red(x) \lor (\blue(x) \land \nextf(\blue(x) \funtil \red(x))) $.}\label{fig:exp_law_counterexample_until_forall}
    \end{figure}

    Consider the \emph{until-forall} case with the formula $\blue(x) \funtil \red(x)$. In the model shown in \autoref{fig:exp_law_counterexample_until_forall} $a_0$ satisfies the expanded formula, since for the atomic counterparts $a_1$ and $b_1$ $\sigma , \set{x \mapsto a_0} \vDash \blue(x) \funtil \red(x)$ and $\sigma , \set{x \mapsto a_0} \vDash \blue(x) \funtil \red(x)$, with the world where all counterparts satisfy $\red(x)$ being reached after two and one steps, respectively. However, we have that $\sigma , \set{x \mapsto a_0} \not \vDash \nextf(\blue(x) \funtil \red(x))$ since there is no single world $\w_n$ where all counterparts after $n$ steps satisfy $\red(x)$.

\pagenumbering{gobble}


\section*{Supplementary material}

## C   Agda example

In this section we illustrate a simple usage of our library in Agda to formalise a counterpart model and prove that some formulae hold on the model, implementing the running example shown in Figure 2.

Our library is able to deal with terms and algebras defined on arbitrary multi-sorted signatures, with directed graphs as a specific instance of this case. To formalise the examples shown in this paper, we begin by defining the signature of graphs Gr, and we show an example graph $G_1$ as algebra on the given signature while omitting the rest of the graphs of the model.

```
import Data.Unit
pattern * = lift Data.Unit.tt

data Gr-Sorts : Set where
  Edge : Gr-Sorts
  Node : Gr-Sorts

data Gr-Functions : Set where
  s : Gr-Functions
  t : Gr-Functions

Gr : Signature
Gr = record { S = Gr-Sorts
            ; F = Gr-Functions
            ; signF = λ { s → [ Edge ] ↦ Node
                        ; t → [ Edge ] ↦ Node
                        }
            } where open import Data.Vec using ([_])

data G₀-Edges : Set where e0 e1 e2 : G₀-Edges
data G₀-Nodes : Set where n0 n1 n2 : G₀-Nodes

G₀ : Σ-Algebra Gr
G₀ = record { S = λ { Edge → G₀-Edges ; Node → G₀-Nodes }
            ; F = λ { s → λ { (e0 , *) → n0
                            ; (e1 , *) → n1
                            ; (e2 , *) → n2
                            }
                    ; t → λ { (e0 , *) → n1
                            ; (e1 , *) → n2
                            ; (e2 , *) → n0
                            }
                    }
            }

-- <omitted definitions of G₁, G₂>
```

Given two graphs, we can capture the notion of graph morphism as given in definition Definition 2.2 by specialising a more general notion of relation between algebras $\Sigma$-Rel in our signature of graphs. We show an example in this case for the relation $C_0$ in Figure 2: note how we require a proof $\rho$-homo stating that the definition given is indeed a relational graph (homo)morphism.

```
data C₀-Edges : G₀-Edges → G₁-Edges → Set where
    e0e4 : C₀-Edges e0 e4
    e1e3 : C₀-Edges e1 e3

data C₀-Nodes : G₀-Nodes → G₁-Nodes → Set where
    n0n4 : C₀-Nodes n0 n4
    n1n3 : C₀-Nodes n1 n3
    n2n4 : C₀-Nodes n2 n4

C₀ : Σ-Rel G₀ G₁
C₀ = record { ρ = λ { {Edge} → C₀-Edges
                    ; {Node} → C₀-Nodes
                    }
            ; ρ-homo = λ { s (e0e4 , *) → n0n4
                         ; s (e1e3 , *) → n1n3
                         ; t (e0e4 , *) → n1n3
                         ; t (e1e3 , *) → n2n4
                         }
            }

data C₁-Edges : G₁-Edges → G₂-Edges → Set where
    e4e5₂ : C₁-Edges e4 e5
data C₁-Nodes : G₁-Nodes → G₂-Nodes → Set where
    n3n5₂ : C₁-Nodes n3 n5
    n4n5₂ : C₁-Nodes n4 n5

-- <omitted definitions of C₁, C₂, C₃>
```

We can now define the set of worlds W and the accessibility relation $\_\leadsto\_$ of the model, mirroring the structure given in Figure 1. Finally, we define our CounterpartModel M by assigning a graph to each world with d and a graph morphism to each accessibility relation with $\mathcal{C}$.

```
data W : Set where
    ω₀ ω₁ ω₂ : W

data _⤳_ : W → W → Set where
    a₀ : ω₀ ⤳ ω₁
    a₁ a₂ : ω₁ ⤳ ω₂
    a₃ : ω₂ ⤳ ω₂

d : W → Σ-Algebra Gr
```

```
d ω₀ = G₀
d ω₁ = G₁
d ω₂ = G₂

𝒞 : ∀ {A B} → A ↝ B → Σ-Rel (d A) (d B)
𝒞 a₀ = C₀
𝒞 a₁ = C₁
𝒞 a₂ = C₂
𝒞 a₃ = C₃

M : CounterpartModel Gr
M = record { W = W ; d = d ; _↝_ = _↝_ ; 𝒞 = 𝒞 }
```

A CounterpartTrace is defined by indicating the accessibility relations of the model, and it is indexed by the first world it begins with; we consider the following examples by defining the trace $\sigma = (C_0, C_1, C_3, C_3, \dots)$ used in Section 3.3.

```
– Repeating trace on ω₂ with a₃
σ-loop-ω₂ : CounterpartTrace M ω₂
σ-loop-ω₂ .B = _
σ-loop-ω₂ .rel = a₃
σ-loop-ω₂ .tail = σ-loop-ω₂

σ : CounterpartTrace M ω₀
σ = a₀ ⇀ a₁ ⇀ σ-loop-ω₂
```

We formalise the shorthand formulae in Section 3.3. We use de Bruijn indices in variables to take care of binders of the quantifiers and ensure well-scopedness of formulae, and we use some helpers v0, v1, v2 to indicate variables in our terms.

```
present : ∀ {τ} → QLTL (τ :: [])
present {τ} = ∃< τ > v1 ≡ᵗ v0

notPresent : ∀ {τ} → QLTL (τ :: [])
notPresent {τ} = ∀< τ > v1 ≢ᵗ v0

nextPreserved : ∀ {τ} → QLTL (τ :: [])
nextPreserved = present ∧ O present

nextDealloc : ∀ {τ} → QLTL (τ :: [])
nextDealloc = present ∧ A notPresent

loop : ∀ {n} {Γ : Ctx n} → QLTL (Edge :: Γ)
loop = s $ (v0 , *) ≡ᵗ t $ (v0 , *)

worldHasLoop : QLTL []
worldHasLoop = ∃< Edge > loop

hasLoop : QLTL (Node :: [])
```

hasLoop = ∃< Edge > (s $ (v0 , *) ≡$^t$ v1 ∧ loop)

adjacent : QLTL (Node :: Node :: [])
adjacent = ∃< Edge > ((s $ (v0 , *) ≡$^t$ v1 ∧ t $ (v0 , *) ≡$^t$ v2)
                              ∨ (s $ (v0 , *) ≡$^t$ v2 ∧ t $ (v0 , *) ≡$^t$ v1))

composable : QLTL (Edge :: Edge :: [])
composable = ∃< Node > (s $ (v2 , *) ≡$^t$ v0 ∧ t $ (v1 , *) ≡$^t$ v0)

haveComposition : QLTL (Edge :: Edge :: [])
haveComposition = composable ∧ ∃< Edge > (s $ (v0 , *) ≡$^t$ s $ (v2 , *)
                                                    ∧ t $ (v0 , *) ≡$^t$ t $ (v1 , *))

willMerge : ∀ {τ} → QLTL (τ :: τ :: [])
willMerge {τ} = v0 ≢$^t$ v1 ∧ ◊ v0 ≡$^t$ v1

alwaysPreserved : ∀ {τ} → QLTL (τ :: [])
alwaysPreserved {τ} = □ present {τ}

willBecomeLoop : QLTL (Edge :: [])
willBecomeLoop = ! loop ∧ ◊ loop

We show how the validity of all example formulae shown in Section 3.3 can be formalised in Agda using our framework. Many of these examples are simple enough that Agda can automatically provide the proof term for the formula.

```
_ : skip 0 σ , (e0 , *) ⊨ nextPreserved {Edge}
_ = (e0 , refl) , ((e4 , *) , (e0e4 , *) , e4 , refl)
_ : skip 0 σ , (n1 , *) ⊨ nextPreserved {Node}
_ = (n1 , refl) , ((n3 , *) , (n1n3 , *) , n3 , refl)
_ : skip 0 σ , (e2 , *) ⊨ ! nextPreserved {Edge}
_ = λ { ((e2 , refl) , ()) }

_ : skip 1 σ , (n3 , *) ⊨ ! nextDealloc {Node}
_ = λ { ((.n3 , refl) , ¬d) → ¬d (n5 , *) (n3n5₂ , *) n5 refl }
_ : skip 1 σ , (e3 , *) ⊨ nextDealloc {Edge}
_ = (e3 , refl) , (λ { (e5 , *) (() , *) e5 refl })
_ : skip 1 σ , (e4 , *) ⊨ ! nextDealloc {Edge}
_ = λ { ((e4 , refl) , ¬d) → ¬d (e5 , *) (e4e5₂ , *) e5 refl }

_ : skip 0 σ , (e0 , *) ⊨ ! loop
_ = λ ()
_ : skip 1 σ , (e3 , *) ⊨ ! loop
_ = λ ()
_ : skip 2 σ , (e5 , *) ⊨ loop
_ = refl

_ : skip 0 σ , (n0 , *) ⊨ ! hasLoop
```

$\_ = \lambda \{ (e0, ()) ; (e1, ()) ; (e2, ()) \}$
$\_ : \text{skip } 1\ \sigma, (n3, *) \vDash !\ \text{hasLoop}$
$\_ = \lambda \{ (e3, ()) ; (e4, ()) \}$
$\_ : \text{skip } 2\ \sigma, (n5, *) \vDash \text{hasLoop}$
$\_ = e5, (\text{refl}, \text{refl})$

$\_ : \text{skip } 0\ \sigma, * \vDash \exists < \text{Node} > (!\ \text{hasLoop})$
$\_ = n0, (\lambda \{ (e0, ()) ; (e1, ()) ; (e2, ()) \})$
$\_ : \text{skip } 1\ \sigma, * \vDash \exists < \text{Node} > (!\ \text{hasLoop})$
$\_ = n3, \lambda \{ (e3, ()) ; (e4, ()) \}$
$\_ : \text{skip } 2\ \sigma, * \vDash \exists < \text{Node} > \text{hasLoop}$
$\_ = n5, (e5, (\text{refl}, \text{refl}))$

$\_ : \text{skip } 0\ \sigma, (n0, n1, *) \vDash \text{adjacent}$
$\_ = e0, \text{inj}_1\ (\text{refl}, \text{refl})$
$\_ : \text{skip } 1\ \sigma, (n3, n4, *) \vDash \text{adjacent}$
$\_ = e3, \text{inj}_1\ (\text{refl}, \text{refl})$
$\_ : \text{skip } 1\ \sigma, (n3, n4, *) \vDash O\ \text{adjacent}$
$\_ = ((n5, n5, *), (n3n5_2, n4n5_2, *), e5, \text{inj}_1\ (\text{refl}, \text{refl}))$

$\_ : \text{skip } 0\ \sigma, (e0, e1, *) \vDash \text{composable}$
$\_ = n1, \text{refl}, \text{refl}$
$\_ : \text{skip } 0\ \sigma, (e0, e1, *) \vDash O\ \text{composable}$
$\_ = (e4, e3, *), (e0e4, e1e3, *), n3, \text{refl}, \text{refl}$
$\_ : \text{skip } 1\ \sigma, (e3, e4, *) \vDash !\ O\ \text{composable}$
$\_ = \lambda \{ () \}$

$\_ : \text{skip } 0\ \sigma, (e0, e1, *) \vDash !\ \text{haveComposition}$
$\_ = \lambda \{ ((n1, \text{refl}, \text{refl}), e2, ()) \}$
$\_ : \text{skip } 0\ \sigma, (e0, e1, *) \vDash !\ O\ \text{haveComposition}$
$\_ = \lambda \{ ((e4, e3, *), (e0e4, e1e3, *), (n3, \text{refl}, \text{refl}), e3, ())$
$\qquad ; ((e4, e3, *), (e0e4, e1e3, *), (n3, \text{refl}, \text{refl}), e4, ()) \}$
$\_ : \text{skip } 2\ \sigma, (e5, e5, *) \vDash O\ \text{haveComposition}$
$\_ = (e5, (e5, *)), (e5e5, e5e5, *), (n5, \text{refl}, \text{refl}), e5, \text{refl}, \text{refl}$

$\_ : \text{skip } 1\ \sigma, * \vDash \exists < \text{Node} > \exists < \text{Node} > \text{willMerge}$
$\_ = n3, (n4, (\lambda ()), 1$
$\qquad , (\lambda \{ \text{zero } (s \leq s\ z \leq n)$
$\qquad\qquad \to (n4, (n3, *)), ((\text{refl}, \text{refl}, *), *)$
$\qquad\qquad ; (\text{suc }\_) (s \leq s\ ()) \})$
$\qquad , (n5, (n5, *)), ((n5, (n4n5_2, \text{refl})), (n5, n3n5_2, \text{refl}), *)$
$\qquad , \text{refl})$

$\_ : \text{skip } 0\ \sigma, * \vDash \exists < \text{Node} > \exists < \text{Node} > \text{willMerge}$
$\_ = n0, (n1, (\lambda ()), 2,$
$\qquad (\lambda \{ \text{zero } (s \leq s\ z \leq n) \to (n1, (n0, *)), ((\text{refl}, \text{refl}, *), *)$

$$; (\text{suc zero}) \ (\text{s}\leq\text{s} \ (\text{s}\leq\text{s z}\leq\text{n}))$$
$$\rightarrow (\text{n3} , \text{n4} , *) , (((\text{n3} , \text{n1n3} , \text{refl})$$
$$, ((\text{n4} , \text{n0n4} , \text{refl}) , *)) , *) \})$$
$$, (\text{n5} , (\text{n5} , *)) , ((\text{n3} , (\text{n1n3} , (\text{n5} , (\text{n3n5}_2 , \text{refl}))))$$
$$, ((\text{n4} , \text{n0n4} , (\text{n5} , \text{n4n5}_2 , \text{refl})) , *)) , \text{refl})$$

\_` : skip 0 $\sigma$ , * ⊨ ! ∀< Edge > alwaysPreserved
\_` ¬∀ with ¬∀ e2
... | inj$_2$ *always* with *always* 1
... | ()

\_ : skip 0 $\sigma$ , * ⊨ ∃< Edge > willBecomeLoop
\_ = e0 , (($\lambda$ ()) , 2 , ($\lambda$ { zero (s≤s z≤n) → (e0 , *) , ((refl , *) , *)
                ; (suc zero) (s≤s (s≤s z≤n)) →
                       (e4 , *) , ((e4 , e0e4 , refl) , *) , *
               })
     , ((e5 , *) , (((e4 , (e0e4 , e5 , (e4e5$_2$ , refl))) , *) , refl)))

\_ : skip 0 $\sigma$ , * ⊨ (∃< Edge > s \$ (v0 , *) $\not\equiv^t$ t \$ (v0 , *))
           U (∃< Node > ∀< Node > v0 $\equiv^t$ v1)
\_ = 2 , ($\lambda$ { zero (s≤s z≤n) → * , (* , (e2 , ($\lambda$ ())))
        ; (suc zero) (s≤s (s≤s z≤n)) → * , (* , (e3 , ($\lambda$ ())))
        }) , * , * , n5 , ($\lambda$ { n5 → refl })

∃□loop : skip 2 $\sigma$ , * ⊨ ∃< Edge > (□ loop)
∃□loop = e5 , inj$_2$ □loop
  where
    □loop : \_
    □loop zero = (e5 , *) , (refl , *) , refl
    □loop (suc $i$) with □loop $i$
    ... | ($e$ , *) , ($\rho$-e5 , *) , *isLoop* = ($e$ , *) , (((e5 , (e5e5 , $\rho$-e5)) , *) , *isLoop*)

\_ : skip 0 $\sigma$ , * ⊨ ∃< Edge > (◊ □ loop)
\_ = e0 , 2 , ($\lambda$ { zero (s≤s z≤n) → (e0 , *) , (refl , *) , *
              ; (suc zero) (s≤s (s≤s z≤n))
                 → (e4 , *) , ((e4 , e0e4 , refl) , *) , * })
     , (e5 , *) , ((e4 , (e0e4 , (e5 , (e4e5$_2$ , refl)))) , *)
     , proj$_2$ ∃□loop



\end{document}

%% file: work.tikzstyles
\tikzstyle{set}=[draw=black, fill=none, shape=ellipse, minimum width=2cm, minimum height=2.25cm]
\tikzstyle{element}=[draw=black, fill={rgb,255: red,0; green,0; blue,0}, shape=circle, minimum size=0.175cm, inner sep=0cm]
\tikzstyle{element_b}=[draw=black, fill={rgb,255: red,54; green,133; blue,236}, shape=circle, minimum size=0.175cm, inner sep=0cm]
\tikzstyle{element_r}=[draw=black, fill={rgb,255: red,230; green,31; blue,31}, shape=circle, minimum size=0.175cm, inner sep=0cm]
\tikzstyle{element_p}=[draw=black, fill={rgb,255: red,181; green,71; blue,141}, shape=circle, minimum size=0.175cm, inner sep=0cm]
\tikzstyle{object}=[fill=none, draw=none, outer sep=0.1cm, inner sep=0cm]
\tikzstyle{category small}=[fill=none, draw=black, shape=rectangle, rounded corners=10, minimum width=9cm, minimum height=2cm]
\tikzstyle{category small mid}=[fill=none, draw=black, shape=rectangle, rounded corners=10, minimum width=9cm, minimum height=3cm]
\tikzstyle{category mid}=[fill=none, draw=black, shape=rectangle, rounded corners=10, minimum width=9cm, minimum height=4.5cm]
\tikzstyle{category midmid}=[fill=none, draw=black, shape=rectangle, rounded corners=10, minimum width=9cm, minimum height=3cm]
\tikzstyle{category big}=[fill=none, draw=black, shape=rectangle, rounded corners=10, minimum width=9cm, minimum height=6cm]
\tikzstyle{category bigbig}=[fill=none, draw=black, shape=rectangle, rounded corners=10, minimum width=9cm, minimum height=6cm]
\tikzstyle{small set}=[fill=none, draw=black, shape=ellipse, minimum width=1.5cm, minimum height=2cm, outer sep=0.1cm]
\tikzstyle{smaller set}=[fill=none, draw=black, shape=ellipse, minimum width=1.6cm, minimum height=1.5cm, outer sep=0.1cm]

\tikzstyle{functor}=[->, dashed]
\tikzstyle{dashing}=[->, dashed]
\tikzstyle{dotting}=[->, dotted]
\tikzstyle{densely_dotting}=[->, densely dotted]
\tikzstyle{arrow}=[->]
\tikzstyle{arrow_morphism}=[->, dotted, line width=1pt, double distance=3pt]
\tikzstyle{red_arrow}=[->, draw=red]
\tikzstyle{none dashing}=[-, dashed]
\tikzstyle{red_arrow_dashing}=[->, draw=red, dashed]
\tikzstyle{red_arrow_dotting}=[->, draw=red, dotted]

\tikzstyle{arrow1}=[->, black!60]
\tikzstyle{arrow2}=[->, draw=blue]
\tikzstyle{arrow3}=[->, thick, dotted, draw=verdino]
\tikzstyle{arrow4}=[->, black!60]

\tikzstyle{rel_grey}=[black!60]

\tikzstyle{arrow1_dash}=[->, dashed, black!60]
\tikzstyle{arrow2_dash}=[->, dashed, draw=blue]
\tikzstyle{arrow3_dash}=[->, thick, dotted, draw=verdino]
\tikzstyle{arrow4_dash}=[->, dashed, black!60]

%% file: work-definitions.tex
\def\A{\mathcal A}
\def\C{\mathcal C}

\def\X{\mathcal X}

\def\W{\mathcal W}
\def\F{\mathcal F} 

\def\false{\mathsf{false}}
\def\true{\mathsf{true}}
\def\G{\varGamma}

\def\fv{\operatorname{fv}}

\def\always{\square}
\def\eventually{\lozenge}

\def\blue{\textcolor{colore_blue}{\textsf{B}}}

\def\red{\textcolor{colore_red}{\textsf{R}}}

\def\w{\omega}

\def\emptymu{\{\,\}}

\def\next{\mathsf O}
\def\nextall{\mathsf A}
\def\then{\mathsf T}
\def\until{\mathsf U}
\def\wuntil{\mathsf W}
\def\funtil{\mathsf F}

\def\nexte{\next}
\def\nextf{\nextall}

\def\alwayse{\always}
\def\eventuallye{\eventually}
\def\alwaysf{\always^*}
\def\eventuallyf{\eventually^*}

\def\nextP{$\next \phi$\xspace}
\def\nexteP{$\nexte \phi$\xspace}
\def\nextfP{$\nextf \phi$\xspace}
\def\alwaysP{$\always \phi$\xspace}
\def\eventuallyP{$\eventually \phi$\xspace}

\def\alwaysfP{$\alwaysf \phi$\xspace}
\def\eventuallyfP{$\eventuallyf \phi$\xspace}
\def\thenP{$\phi_1 \then \phi_2$\xspace}
\def\untilP{$\phi_1 \until \phi_2$\xspace}
\def\wuntilP{$\phi_1 \wuntil \phi_2$\xspace}
\def\funtilP{$\phi_1 \funtil \phi_2$\xspace}

\def\CTL{\textsf{CTL}\xspace}
\def\LTL{\textsf{LTL}\xspace}
\def\QLTL{\textsf{QLTL}\xspace}
\def\PNF{\textsf{PNF}\xspace}

%% file: symbols.tex
\newunicodechar{′}{\ensuremath{\,'}} 
\newunicodechar{∷}{\ensuremath{::}}
\newunicodechar{∙}{\ensuremath{\bullet}}
\newunicodechar{∎}{\ensuremath{\mathrel{\mathsmaller{\mathsmaller{\blacksquare}}}}} 

\newunicodechar{₀}{\ensuremath{_0}}
\newunicodechar{₁}{\ensuremath{_1}}
\newunicodechar{₂}{\ensuremath{_2}}
\newunicodechar{₃}{\ensuremath{_3}}

\newunicodechar{ₗ}{\ensuremath{_l}}
\newunicodechar{ᵣ}{\ensuremath{_r}}
\newunicodechar{ˡ}{\ensuremath{^l}}
\newunicodechar{ʳ}{\ensuremath{^r}}
\newunicodechar{ᵣ}{\ensuremath{_r}}
\newunicodechar{ᵢ}{\ensuremath{_i}}

\newunicodechar{≤}{\ensuremath{\le}}
\newunicodechar{≥}{\ensuremath{\ge}}

\newunicodechar{→}{\ensuremath{\rightarrow}}
\newunicodechar{⇒}{\ensuremath{\Rightarrow}}

\newunicodechar{∋}{\ensuremath{\ni}}
\newunicodechar{∀}{\ensuremath{\forall}}
\newunicodechar{∃}{\ensuremath{\exists}}
\newunicodechar{⊎}{\ensuremath{\uplus}}
\newunicodechar{⊢}{\ensuremath{\vdash}}
\newunicodechar{×}{\ensuremath{\times}}
\newunicodechar{⟨}{\ensuremath{\langle}}
\newunicodechar{⟩}{\ensuremath{\rangle}}
\newunicodechar{ℕ}{\ensuremath{\mathbb{N}}}
\newunicodechar{∘}{\ensuremath{\circ}}
\newunicodechar{⟦}{\ensuremath{\llbracket}}
\newunicodechar{⟧}{\ensuremath{\rrbracket}}
\newunicodechar{∈}{\ensuremath{\in}}
\newunicodechar{⊔}{\ensuremath{\sqcup}}

\newunicodechar{ℓ}{\ensuremath{\ell}}

\newunicodechar{Δ}{\ensuremath{\Delta}}
\newunicodechar{Γ}{\ensuremath{\Gamma}}
\newunicodechar{Σ}{\ensuremath{\Sigma}}

\newunicodechar{α}{\ensuremath{\alpha}}
\newunicodechar{β}{\ensuremath{\beta}}
\newunicodechar{τ}{\ensuremath{\tau}}
\newunicodechar{Π}{\ensuremath{\Pi}}

\newunicodechar{ρ}{\ensuremath{\rho}}
\newunicodechar{σ}{\ensuremath{\sigma}}
\newunicodechar{λ}{\ensuremath{\lambda}}
\newunicodechar{ξ}{\ensuremath{\xi}}
\newunicodechar{θ}{\ensuremath{\theta}}

\newunicodechar{η}{\ensuremath{\eta}}

\newunicodechar{⟪}{\ensuremath{\mathrel{\mathsmaller{\ll}}}}
\newunicodechar{⟫}{\ensuremath{\mathrel{\mathsmaller{\gg}}}}
\newunicodechar{⨟}{\ensuremath{\fatsemi}}
\newunicodechar{〕}{\ensuremath{\rrbracket}}
\newunicodechar{〔}{\ensuremath{\llbracket}}

\newunicodechar{ƛ}{\textcrlambda}

\newunicodechar{⟶}{\ensuremath{\rightarrow}} 
\newunicodechar{↠}{\ensuremath{\twoheadrightarrow}} 
\newunicodechar{⇉}{\ensuremath{\rightrightarrows}}

\newunicodechar{ᵐ}{\ensuremath{^m}}
\newunicodechar{ⁿ}{\ensuremath{^n}}
\newunicodechar{ᵏ}{\ensuremath{^ᵏ}}

\newunicodechar{ˢ}{\ensuremath{^s}}
\newunicodechar{ₛ}{\ensuremath{_s}}

\newunicodechar{⃰}{\ensuremath{^*}}

\newunicodechar{⁽}{\ensuremath{^(}}
\newunicodechar{⁾}{\ensuremath{^)}}

\newunicodechar{⊔}{\ensuremath{\sqcup}}

\newunicodechar{▻}{\ensuremath{\triangleright}}
\newunicodechar{◅}{\ensuremath{\triangleleft}}

\newunicodechar{ε}{\ensuremath{\varepsilon}}

\newunicodechar{𝓕}{\ensuremath{\mathcal F}}
\newunicodechar{𝓢}{\ensuremath{\mathcal S}}

\newunicodechar{𝑓}{\ensuremath{\mathcal F}}
\newunicodechar{⇝}{\ensuremath{\rightsquigarrow}}

\newunicodechar{◯}{\ensuremath{\mathsf O}}
\newunicodechar{𝑇}{\ensuremath{T}}

\newunicodechar{⇀}{\ensuremath{\rightharpoonup}}

\newunicodechar{◇}{\ensuremath{\eventually}}
\newunicodechar{□}{\ensuremath{\always}}

\newunicodechar{≡}{\ensuremath{\equiv}}
\newunicodechar{≢}{\ensuremath{\not \equiv}}

\newunicodechar{⊨}{\ensuremath{\vDash}}

\newunicodechar{𝓒}{\ensuremath{\mathcal C}}

\newunicodechar{ω}{\ensuremath{\omega}}

\newunicodechar{∧}{\ensuremath{\land}}
\newunicodechar{∨}{\ensuremath{\lor}}

\newunicodechar{¬}{\ensuremath{\neg}}

\newunicodechar{↦}{\ensuremath{\mapsto}}

\newunicodechar{ᵗ}{\ensuremath{^t}}

%% file: algebraic_counterpart_w.tikz
\begin{tikzpicture}
	\begin{pgfonlayer}{nodelayer}
		\node [style=object] (0) at (-4.5, 0) {$\omega_0$};
		\node [style=none] (2) at (-2.5, 0.75) {$C_0$};
		\node [style=object] (3) at (-0.5, 0) {$\omega_1$};
		\node [style=none] (5) at (1.5, -1.25) {$C_2$};
		\node [style=object] (6) at (3.5, 0) {$\omega_2$};
		\node [style=none] (7) at (1.5, 1.25) {$C_1$};
		\node [style=none] (8) at (6.25, 0) {$C_3$};
	\end{pgfonlayer}
	\begin{pgfonlayer}{edgelayer}
		\draw [style=arrow1, in=180, out=0] (0) to (3);
		\draw [style=arrow2_dash, bend left=15] (3) to (6);
		\draw [style=arrow3, in=195, out=-15] (3) to (6);
		\draw [style=arrow4, in=45, out=-30, loop] (6) to ();
	\end{pgfonlayer}
\end{tikzpicture}

%% file: running_example_graph.tikz
\begin{tikzpicture}[shorten > = 2pt, shorten < = 2pt, auto, node distance = 2cm, semithick]
    \tikzstyle{vertex} = [circle, draw = black, thick, fill = white, minimum size = 2mm]
    \node[vertex] (n0) [] {$n_0$};
    \node[vertex] (n1) [above right = 0.5cm and 1.2cm of n0] {$n_1$};
    \node[vertex] (n2) [below right = 0.5cm and 1.2cm of n0] {$n_2$};
    \node[vertex] (n4) [right = 3cm of n2] {$n_4$};
    \node[vertex] (n3) [right = 3cm of n1] {$n_3$};

    \path[->] (n4) edge [bend right] node [swap] (e4) {$e_4$} (n3);
    \path[->] (n3) edge [bend right] node [swap] (e3) {$e_3$} (n4);
    \path[->] (n0) edge [bend left]  node        (e0) {$e_0$} (n1);
    \path[->] (n1) edge [bend left]  node        (e1) {$e_1$} (n2);
    \path[->] (n2) edge [bend left]  node        (e2) {$e_2$} (n0);

    \node[vertex] (n5) [right = 3cm of e4] {$n_5$};
    \path[->] (n5) edge [loop left, looseness=10, out=150, in=-150] node (e5) {$e_5$} (n5);

    \path[style=arrow1_dash]                                     (n1) edge [bend left] (n3);
    \path[style=arrow1_dash]                                     (n2) edge [bend right] (n4);
    \path[style=arrow1_dash]                                     (n0) edge [bend right=60, out=270, in=240] (n4);
    \path[style=arrow1_dash, shorten < =  6pt, shorten > =  5pt] (e0) edge [bend right, out=-38] (e4);
    \path[style=arrow1_dash, shorten < = -2pt, shorten > = -2pt] (e1) edge [bend left]  (e3);

    \path[style=arrow3_dash, shorten < =  4pt, shorten > = -2pt, bend left=22pt] (e3) edge [bend left]  (e5);
    \path[style=arrow2_dash, shorten < = -2pt, shorten > = -3pt] (e4) edge [bend right] (e5);

    \path[style=arrow3_dash] (n3) edge [bend left] (n5);
    \path[style=arrow2_dash, shorten < =  1pt, shorten > = -1pt] (n3) edge [bend left, transform canvas={yshift=0.7ex}] (n5);

    \path[style=arrow3_dash] (n4) edge [bend right] (n5);
    \path[style=arrow2_dash, shorten < =  1pt, shorten > = -1pt] (n4) edge [bend right, transform canvas={yshift=-0.7ex}] (n5);

    \path[style=arrow4_dash] (n5) edge [loop, looseness=6, in=65, out=115]  node [swap] (c1) {} (n5);
    \path[style=arrow4_dash] (e5) edge [loop, shorten > = -1pt,  shorten < = -1pt, looseness=10, in=60, out=120] node [swap] (c2) {} (e5);
\end{tikzpicture}

%% file: running_example_graph2.tikz
\begin{tikzpicture}[shorten > = 2pt, shorten < = 2pt, auto, node distance = 2cm, semithick]15
    \tikzstyle{vertex} = [circle, draw = black, thick, fill = white, minimum size = 2mm]

    \node[vertex] (n1) [] {$n_1$};
    \node[vertex] (n2) [below = 1cm  of n1] {$n_2$};

    \node[vertex] (n0) [below left = 0.4cm and 2.8cm of n1] {$n_0$};

    \node[vertex] (n3) [above right = 0.4cm and 2.8cm of n1] {$n_3$};
    \node[vertex] (n4) [below = 1cm   of n3] {$n_4$};
    \node[vertex] (n5) [below = 1cm   of n4] {$n_5$};

    \node[vertex] (n6) [below right = 0.4cm and 2.8cm of n3] {$n_6$};

    \path[->] (n1) edge [looseness=7, out=-90-150, in=-90+150] node [] (e0) {$e_0$} (n1);
    \path[->] (n1) edge []                             node [] (e1_) {$e_1$} (n2);
    \path[->] (n2) edge [looseness=7, out=90-150, in=90+150]   node [] (e2_) {$e_2$} (n2);

    \path[->] (n3) edge [] node [] (e3) {$e_4$} (n4);
    \path[->] (n4) edge [] node [] (e4) {$e_5$} (n5);
    \path[->] (n3) edge [looseness=7, out=90+90+90-150, in=90+90+90+150] node [] (e5) {$e_3$} (n3);
    \path[->] (n5) edge [looseness=7, out=90-150, in=90+150] node [] (e6) {$e_6$} (n5);

    \path[->] (n6) edge [looseness=5, out=90-150, in=90+150] node [] (e7) {$e_7$} (n6);

    \path[style=arrow1_dash] (n0) edge [] (n1);
    \path[style=arrow1_dash] (n0) edge [] (n2);

    \path[style=arrow1_dash] (n1) edge [] (n3);
    \path[style=arrow1_dash] (n1) edge [] (n4);
    \path[style=arrow1_dash] (n2) edge [] (n4);
    \path[style=arrow1_dash] (n2) edge [] (n5);

    \path[style=arrow1_dash] (e1_) edge  [shorten > = 5pt,  shorten < = -1pt, swap] (e3);
    \path[style=arrow1_dash] (e1_) edge  [shorten > = 5pt,  shorten < = -1pt, swap] (e4);

    \path[style=arrow1_dash] (e3) edge  [shorten > = -1pt,  shorten < = -1pt, swap] (e7);

    \path[style=arrow1_dash] (n3) edge [] (n6);
    \path[style=arrow1_dash] (n4) edge [] (n6);

    \path[style=arrow1_dash] (n6) edge [looseness=10, out=180-150, in=180+150] (n6);
    \path[style=arrow1_dash] (e7) edge [looseness=10, shorten > = -3pt,  shorten < = -1pt, out=180-150, in=180+150]  (e7);

\end{tikzpicture}

%% file: then_counterexample.tikz
\begin{tikzpicture}
	\begin{pgfonlayer}{nodelayer}
		\node [style={element_r}] (12) at (10, 0.75) {};
		\node [style={element_b}] (9) at (5, 0) {};
		\node [style=none] (17) at (0, -3) {$\omega_0$};
		\node [style=none] (19) at (5, -3) {$\omega_1$};
		\node [style=none] (22) at (10, -3) {$\omega_2$};
		\node [style=none] (22) at (2.5, -2) {$C_0$};
		\node [style=none] (23) at (7.5, -2) {$C_1$};
		\node [style=none] (64) at (0, -0.75) {$a_0$};
		\node [style=none] (69) at (5, -0.75) {$a_1$};
		\node [style=none] (71) at (10.75, 0.75) {$a_2$};
		\node [style=none] (72) at (10.75, -0.75) {$b_2$};
		\node [style={element_b}] (0) at (0, 0) {};
		\node [style=set] (103) at (5, 0) {};
		\node [style={element_b}] (105) at (10, -0.75) {};
		\node [style=set] (108) at (10, 0) {};
		\node [style=set] (109) at (0, 0) {};
		\node [style=none] (110) at (12.5, -2) {$C_2$};
	\end{pgfonlayer}
	\begin{pgfonlayer}{edgelayer}
		\draw [style=arrow1] (0) to (9);
		\draw [style=arrow1] (9) to (12);
		\draw [style=arrow1] (9) to (105);
	\end{pgfonlayer}
\end{tikzpicture}

%% file: exp_law_counterexample_until.tikz
\begin{tikzpicture}
	\begin{pgfonlayer}{nodelayer}
		\node [style={element_r}] (12) at (10, 0) {};
		\node [style=element] (9) at (5, 0.75) {};
		\node [style=none] (17) at (0, -3) {$\omega_0$};
		\node [style=none] (18) at (5, -3) {};
		\node [style=none] (19) at (5, -3) {$\omega_1$};
		\node [style=none] (20) at (5, -3) {};
		\node [style=none] (22) at (10, -3) {$\omega_2$};
		\node [style=none] (22) at (2.5, -2) {$C_0$};
		\node [style=none] (23) at (7.5, -2) {$C_1$};
		\node [style=none] (64) at (0, -0.75) {$a_0$};
		\node [style=none] (69) at (5.5, 1.5) {$a_1$};
		\node [style=none] (71) at (10.75, 0) {$a_2$};
		\node [style={element_b}] (0) at (0, 0) {};
		\node [style=set] (103) at (5, 0) {};
		\node [style=set] (108) at (10, 0) {};
		\node [style=set] (109) at (0, 0) {};
		\node [style=none] (110) at (12.5, -2) {$C_2$};
		\node [style={element_b}] (111) at (5, -0.75) {};
		\node [style=none] (112) at (5.5, -1.5) {$b_1$};
	\end{pgfonlayer}
	\begin{pgfonlayer}{edgelayer}
		\draw [style=arrow1] (0) to (9);
		\draw [style=arrow1] (9) to (12);
		\draw [style=arrow1] (0) to (111);
	\end{pgfonlayer}
\end{tikzpicture}

%% file: exp_law_counterexample_until_forall.tikz
\begin{tikzpicture}
	\begin{pgfonlayer}{nodelayer}
		\node [style={element_r}] (12) at (10, 0.75) {};
		\node [style={element_b}] (9) at (5, 0.75) {};
		\node [style=none] (17) at (0, -3) {$\omega_0$};
		\node [style=none] (18) at (5, -3) {};
		\node [style=none] (19) at (5, -3) {$\omega_1$};
		\node [style=none] (20) at (5, -3) {};
		\node [style=none] (22) at (10, -3) {$\omega_2$};
		\node [style=none] (22) at (2.5, -2) {$C_0$};
		\node [style=none] (23) at (7.5, -2) {$C_1$};
		\node [style=none] (64) at (0, -0.75) {$a_0$};
		\node [style=none] (69) at (5.5, 1.5) {$a_1$};
		\node [style=none] (71) at (10.5, 1.5) {$a_2$};
		\node [style={element_b}] (0) at (0, 0) {};
		\node [style=set] (103) at (5, 0) {};
		\node [style=set] (108) at (10, 0) {};
		\node [style=set] (109) at (0, 0) {};
		\node [style=none] (110) at (12.5, -2) {$C_2$};
		\node [style={element_r}] (111) at (5, -0.75) {};
		\node [style=none] (112) at (5.5, -1.5) {$b_1$};
		\node [style=element] (113) at (10, -0.75) {};
		\node [style=none] (114) at (10.5, -1.5) {$b_2$};
	\end{pgfonlayer}
	\begin{pgfonlayer}{edgelayer}
		\draw [style=arrow1] (0) to (9);
		\draw [style=arrow1] (0) to (111);
		\draw [style=arrow1] (111) to (113);
		\draw [style=arrow1] (9) to (12);
	\end{pgfonlayer}
\end{tikzpicture}